%Paper: 9109046
%From: Andrea Pasquinucci <pasquinu@phoenix.Princeton.EDU>
%Date: Tue, 24 Sep 91 18:32:24 EDT

%--------------------------------------------------------------------
%
% HERMITIAN vs. ANTI-HERMITIAN 1-MATRIX MODELS AND THEIR HIERARCHIES
%
%--------------------------------------------------------------------
%
%                                      Andrea PASQUINUCCI, 1988
%          MACROS (panda.tex)          S.I.S.S.A., Trieste, Italy
%                                      (Revised 1991, Princeton, USA)
%
%--------------------------------------------------------------------
%
%    This are TEX macros. They work with PLAIN TEX (the basis
%    version of TEX). The only problem can be with the double-page
%    format since it depends on the type of postscript and postscript
%    laserwriters you use.
%
%--------------------------------------------------------------------
%
\def\standardrisposta{s }\def\reducedrisposta{r }
\def\doublerisposta{d }\def\cartarisposta{e }\def\amsrisposta{y }
\newcount\ingrandimento \newcount\sinnota \newcount\dimnota
\newcount\unoduecol \newdimen\collhsize \newdimen\tothsize
\newdimen\fullhsize \newcount\controllorisposta
\global\controllorisposta=0
\message{ ********    Welcome to PANDA macros (Plain TeX, AP, 1991)}
\message{ ******** }
\message{       You'll have to answer a few questions in lowercase.}
\message{>  Do you want it in double-page (d), reduced (r)}
\message{or standard format (s) ? }\read-1 to\risposta
\message{>  Do you want it in USA A4 (u) or EUROPEAN A4 (e)}
\message{paper size ? }\read-1 to\srisposta
\message{>  Do you have AMS (math) fonts (y/n) ? }\read-1 to\arisposta
\ifx\risposta\standardrisposta
\message{>> This will come out UNREDUCED << }
\ingrandimento=1200 \dimnota=2 \unoduecol=1 \baselineskip=16pt
\global\controllorisposta=1 \fi
\ifx\risposta\reducedrisposta
\ingrandimento=1095 \dimnota=1 \unoduecol=1 \baselineskip=14pt
\message{>> This will come out REDUCED << }
\global\controllorisposta=1 \fi
\ifx\risposta\doublerisposta
\ingrandimento=1000 \dimnota=2 \unoduecol=2 \baselineskip=12pt
\message{>> You must print this in LANDSCAPE orientation << }
\global\controllorisposta=1 \fi
\ifnum\controllorisposta=0
\message{>>> ERROR IN INPUT, I ASSUME STANDARD UNREDUCED FORMAT <<< }
\ingrandimento=1200 \dimnota=2 \unoduecol=1 \baselineskip=16pt\fi
\def\magnificazione{\magnification=\ingrandimento}  \magnificazione
\ifx\risposta\doublerisposta
\ifx\srisposta\cartarisposta
\tothsize=25.5truecm \collhsize=12.0truecm \vsize=17.0truecm \else
\tothsize=9.4truein \collhsize=4.4truein \vsize=6.8truein \fi \else
\ifx\srisposta\cartarisposta
\tothsize=6.5truein \vsize=24truecm \else
\tothsize=6.5truein \vsize=8.9truein \fi
\collhsize=4.4truein \fi
\tolerance=10000
\leftskip=0pt \rightskip=0pt \sinnota=1 \parskip 0pt plus 2pt
%
%--------------------------------------------------------------------
%
\font\ninerm=cmr9  \font\eightrm=cmr8  \font\sixrm=cmr6
\font\ninei=cmmi9  \font\eighti=cmmi8  \font\sixi=cmmi6
\font\ninesy=cmsy9  \font\eightsy=cmsy8  \font\sixsy=cmsy6
\font\ninebf=cmbx9  \font\eightbf=cmbx8  \font\sixbf=cmbx6
\font\ninett=cmtt9  \font\eighttt=cmtt8 \font\nineit=cmti9
\font\eightit=cmti8 \font\ninesl=cmsl9  \font\eightsl=cmsl8
\skewchar\ninei='177 \skewchar\eighti='177 \skewchar\sixi='177
\skewchar\ninesy='60 \skewchar\eightsy='60 \skewchar\sixsy='60
\hyphenchar\ninett=-1 \hyphenchar\eighttt=-1 \hyphenchar\tentt=-1
\font\tencaps=cmcsc10
\ifx\arisposta\amsrisposta
\font\mmmath=msym10      
\else \gdef\mmmath{\bf}  \fi
\ifnum\ingrandimento=1095 
 
\font\bfone=cmbx10 at 10.95pt

\font\capsone=cmcsc10 at 10.95pt 

\else  
 
\font\bfone=cmbx10 at 12pt

\font\capsone=cmcsc10 at 12pt 
\fi
\def\ttaarr{\bf}   \def\ppaarr{\sl}
\catcode`@=11
\newskip\ttglue
\gdef\tenpoint{\def\rm{\fam0\tenrm}
  \textfont0=\tenrm \scriptfont0=\sevenrm \scriptscriptfont0=\fiverm
  \textfont1=\teni \scriptfont1=\seveni \scriptscriptfont1=\fivei
  \textfont2=\tensy \scriptfont2=\sevensy \scriptscriptfont2=\fivesy
  \textfont3=\tenex \scriptfont3=\tenex \scriptscriptfont3=\tenex
  \textfont\itfam=\tenit \def\it{\fam\itfam\tenit}
  \textfont\slfam=\tensl \def\sl{\fam\slfam\tensl}
  \textfont\ttfam=\tentt \def\tt{\fam\ttfam\tentt}
  \textfont\bffam=\tenbf \scriptfont\bffam=\sevenbf
  \scriptscriptfont\bffam=\fivebf \def\bf{\fam\bffam\tenbf}
  \tt \ttglue=.5em plus.25em minus.15em
  \normalbaselineskip=12pt
  \setbox\strutbox=\hbox{\vrule height8.5pt depth3.5pt width0pt}
  \let\sc=\eightrm \let\big=\tenbig \normalbaselines\rm}
\gdef\ninepoint{\def\rm{\fam0\ninerm}
  \textfont0=\ninerm \scriptfont0=\sixrm \scriptscriptfont0=\fiverm
  \textfont1=\ninei \scriptfont1=\sixi \scriptscriptfont1=\fivei
  \textfont2=\ninesy \scriptfont2=\sixsy \scriptscriptfont2=\fivesy
  \textfont3=\tenex \scriptfont3=\tenex \scriptscriptfont3=\tenex
  \textfont\itfam=\nineit \def\it{\fam\itfam\nineit}
  \textfont\slfam=\ninesl \def\sl{\fam\slfam\ninesl}
  \textfont\ttfam=\ninett \def\tt{\fam\ttfam\ninett}
  \textfont\bffam=\ninebf \scriptfont\bffam=\sixbf
  \scriptscriptfont\bffam=\fivebf \def\bf{\fam\bffam\ninebf}
  \tt \ttglue=.5em plus.25em minus.15em
  \normalbaselineskip=11pt
  \setbox\strutbox=\hbox{\vrule height8pt depth3pt width0pt}
  \let\sc=\sevenrm \let\big=\ninebig \normalbaselines\rm}
\gdef\eightpoint{\def\rm{\fam0\eightrm}
  \textfont0=\eightrm \scriptfont0=\sixrm \scriptscriptfont0=\fiverm
  \textfont1=\eighti \scriptfont1=\sixi \scriptscriptfont1=\fivei
  \textfont2=\eightsy \scriptfont2=\sixsy \scriptscriptfont2=\fivesy
  \textfont3=\tenex \scriptfont3=\tenex \scriptscriptfont3=\tenex
  \textfont\itfam=\eightit \def\it{\fam\itfam\eightit}
  \textfont\slfam=\eightsl \def\sl{\fam\slfam\eightsl}
  \textfont\ttfam=\eighttt \def\tt{\fam\ttfam\eighttt}
  \textfont\bffam=\eightbf \scriptfont\bffam=\sixbf
  \scriptscriptfont\bffam=\fivebf \def\bf{\fam\bffam\eightbf}
  \tt \ttglue=.5em plus.25em minus.15em
  \normalbaselineskip=9pt
  \setbox\strutbox=\hbox{\vrule height7pt depth2pt width0pt}
  \let\sc=\sixrm \let\big=\eightbig \normalbaselines\rm}
\gdef\tenbig#1{{\hbox{$\left#1\vbox to8.5pt{}\right.\n@space$}}}
\gdef\ninebig#1{{\hbox{$\textfont0=\tenrm\textfont2=\tensy
   \left#1\vbox to7.25pt{}\right.\n@space$}}}
\gdef\eightbig#1{{\hbox{$\textfont0=\ninerm\textfont2=\ninesy
   \left#1\vbox to6.5pt{}\right.\n@space$}}}
 %for 10-pt math in 9-pt territory
%
\newbox\fotlinebb \newbox\hedlinebb \newbox\leftcolumn
\gdef\makeheadline{\vbox to 0pt{\vskip-22.5pt
     \fullline{\vbox to8.5pt{}\the\headline}\vss}\nointerlineskip}
\gdef\makehedlinebb{\vbox to 0pt{\vskip-22.5pt
     \fullline{\vbox to8.5pt{}\copy\hedlinebb\hfil
     \line{\hfill\the\headline\hfill}}\vss}
     \nointerlineskip}
\gdef\makefootline{\baselineskip=24pt \fullline{\the\footline}}
\gdef\makefotlinebb{\baselineskip=24pt
    \fullline{\copy\fotlinebb\hfil\line{\hfill\the\footline\hfill}}}
\gdef\doubleformat{\shipout\vbox{\makehedlinebb
     \fullline{\box\leftcolumn\hfil\columnbox}\makefotlinebb}
     \advancepageno}

\gdef\columnbox{\leftline{\pagebody}}
\gdef\line#1{\hbox to\hsize{\hskip\leftskip#1\hskip\rightskip}}
\gdef\fullline#1{\hbox to\fullhsize{\hskip\leftskip{#1}%
\hskip\rightskip}}
\gdef\footnote#1{\let\@sf=\empty
         \ifhmode\edef\#sf{\spacefactor=\the\spacefactor}\/\fi
         #1\@sf\vfootnote{#1}}
\gdef\vfootnote#1{\insert\footins\bgroup
         \ifnum\dimnota=1  \eightpoint\fi
         \ifnum\dimnota=2  \ninepoint\fi
         \ifnum\dimnota=0  \tenpoint\fi
         \interlinepenalty=\interfootnotelinepenalty
         \splittopskip=\ht\strutbox
         \splitmaxdepth=\dp\strutbox \floatingpenalty=20000
         \leftskip=\oldssposta \rightskip=\olddsposta
         \spaceskip=0pt \xspaceskip=0pt
         \ifnum\sinnota=0   \textindent{#1}\fi
         \ifnum\sinnota=1   \item{#1}\fi
         \footstrut\futurelet\next\fo@t}
\gdef\fo@t{\ifcat\bgroup\noexpand\next \let\next\f@@t
             \else\let\next\f@t\fi \next}
\gdef\f@@t{\bgroup\aftergroup\@foot\let\next}
\gdef\f@t#1{#1\@foot}
\gdef\@foot{\strut\egroup}
\gdef\footstrut{\vbox to\splittopskip{}}
\skip\footins=\bigskipamount
\count\footins=1000  \dimen\footins=8in
\catcode`@=12
\tenpoint
\newskip\olddsposta \newskip\oldssposta
\global\oldssposta=\leftskip \global\olddsposta=\rightskip
\gdef\yespagenumbers{\footline={\hss\tenrm\folio\hss}}
\gdef\ciao{\par\vfill\supereject
      \ifnum\unoduecol=2 \if R\lrcol \null\vfill\eject \fi\fi \end}
\ifnum\unoduecol=1 \hsize=\tothsize   \fullhsize=\tothsize \fi
\ifnum\unoduecol=2 \hsize=\collhsize  \fullhsize=\tothsize \fi
\global\let\lrcol=L
\ifnum\unoduecol=1 \output{\plainoutput}\fi
\ifnum\unoduecol=2 \output{\if L\lrcol
       \global\setbox\leftcolumn=\columnbox
       \global\setbox\fotlinebb=\line{\hfill\the\footline\hfill}
       \global\setbox\hedlinebb=\line{\hfill\the\headline\hfill}
       \advancepageno
      \global\let\lrcol=R \else \doubleformat \global\let\lrcol=L \fi
       \ifnum\outputpenalty>-20000 \else\dosupereject\fi}\fi
\def\ifdoublepage{\ifnum\unoduecol=2 }

\def\newline{\hfil\break}
\def\jump{\vskip\baselineskip} \newskip\iinnffrr
\def\sjump{\iinnffrr=\baselineskip
          \divide\iinnffrr by 2 \vskip\iinnffrr}
\def\bjump{\vskip\baselineskip \vskip\baselineskip}
\newcount\nmbnota  \def\clearnmbnota{\global\nmbnota=0}
\def\note#1{\global\advance\nmbnota by 1
    \footnote{$^{\the\nmbnota}$}{#1}}  \clearnmbnota
\def\setnote#1{\expandafter\xdef\csname#1\endcsname{\the\nmbnota}}
\def\formula{$$} \def\endformula{\eqno\numero $$}
\def\fr{\formula} \def\efr{\endformula}
\newcount\frmcount \def\clearfrmcount{\global\frmcount=0}
\def\numero{\global\advance\frmcount by 1   \ifnum\indappcount=0
  {\ifnum\cpcount <1 {\hbox{\rm (\the\frmcount )}}  \else
  {\hbox{\rm (\the\cpcount .\the\frmcount )}} \fi}  \else
  {\hbox{\rm (\applett .\the\frmcount )}} \fi}
\def\nameformula#1{\global\advance\frmcount by 1%
\ifnum\draftnum=0  {\ifnum\indappcount=0%
{\ifnum\cpcount<1\xdef\spzzttrra{(\the\frmcount )}%
\else\xdef\spzzttrra{(\the\cpcount .\the\frmcount )}\fi}%
\else\xdef\spzzttrra{(\applett .\the\frmcount )}\fi}%
\else\xdef\spzzttrra{(#1)}\fi%
\expandafter\xdef\csname#1\endcsname{\spzzttrra}
\eqno \ifnum\draftnum=0 {\ifnum\indappcount=0
  {\ifnum\cpcount <1 {\hbox{\rm (\the\frmcount )}}  \else
  {\hbox{\rm (\the\cpcount .\the\frmcount )}}\fi}   \else
  {\hbox{\rm (\applett .\the\frmcount )}} \fi} \else (#1) \fi $$}
\def\nfr{\nameformula}    
\def\nameali#1{\global\advance\frmcount by 1%
\ifnum\draftnum=0  {\ifnum\indappcount=0%
{\ifnum\cpcount<1\xdef\spzzttrra{(\the\frmcount )}%
\else\xdef\spzzttrra{(\the\cpcount .\the\frmcount )}\fi}%
\else\xdef\spzzttrra{(\applett .\the\frmcount )}\fi}%
\else\xdef\spzzttrra{(#1)}\fi%
\expandafter\xdef\csname#1\endcsname{\spzzttrra}
\eqno \ifnum\draftnum=0  {\ifnum\indappcount=0
  {\ifnum\cpcount <1 {\hbox{\rm (\the\frmcount )}}  \else
  {\hbox{\rm (\the\cpcount .\the\frmcount )}}\fi}   \else
  {\hbox{\rm (\applett .\the\frmcount )}} \fi} \else (#1) \fi}
\clearfrmcount
\newcount\cpcount \def\clearcpcount{\global\cpcount=0}
\newcount\subcpcount \def\clearsubcpcount{\global\subcpcount=0}
\newcount\appcount \def\clearappcount{\global\appcount=0}
\newcount\indappcount \def\clearindappcount{\indappcount=0}
\newcount\sottoparcount 

\def\applett{\ifcase\appcount  \or {A}\or {B}\or {C}\or
{D}\or {E}\or {F}\or {G}\or {H}\or {I}\or {J}\or {K}\or {L}\or
{M}\or {N}\or {O}\or {P}\or {Q}\or {R}\or {S}\or {T}\or {U}\or
{V}\or {W}\or {X}\or {Y}\or {Z}\fi
             \ifnum\appcount<0
    \message{>>  ERROR: counter \appcount out of range <<}\fi
             \ifnum\appcount>26
   \message{>>  ERROR: counter \appcount out of range <<}\fi}
\clearappcount  \clearindappcount
\newcount\connttrre  \def\clearconnttrre{\global\connttrre=0}
\newcount\countref  \def\clearcountref{\global\countref=0}
\clearcountref
\def\chapter#1{\global\advance\cpcount by 1 \clearfrmcount
                 \goodbreak\null\jump\nobreak
                 \clearsubcpcount\clearindappcount
                 \itemitem{\ttaarr\the\cpcount .\qquad}{\ttaarr #1}
                 \par\nobreak\jump\sjump\nobreak}
\def\section#1{\global\advance\subcpcount by 1 \goodbreak\null
                  \sjump\nobreak\ifnum\indappcount=0
                 {\ifnum\cpcount=0 {\itemitem{\ppaarr
               .\the\subcpcount\quad\enskip\ }{\ppaarr #1}\par} \else
                 {\itemitem{\ppaarr\the\cpcount .\the\subcpcount\quad
                  \enskip\ }{\ppaarr #1} \par}  \fi}
                \else{\itemitem{\ppaarr\applett .\the\subcpcount\quad
                 \enskip\ }{\ppaarr #1}\par}\fi\nobreak\jump\nobreak}
\clearsubcpcount
\def\appendix#1{\global\advance\appcount by 1 \clearfrmcount
                  \goodbreak\null\jump\nobreak
                  \global\advance\indappcount by 1 \clearsubcpcount
                  \itemitem{\ttaarr App.\applett\ }{\ttaarr #1}
                  \nobreak\jump\sjump\nobreak}
\clearappcount \clearindappcount
\def\references{\goodbreak\null\jump\nobreak
   \itemitem{}{\ttaarr References} \nobreak\jump\sjump\nobreak}
\clearcountref
\def\introduction{\clearindappcount\clearappcount\clearcpcount
                  \clearsubcpcount\goodbreak\null\jump\nobreak
  \itemitem{}{\ttaarr Introduction} \nobreak\jump\sjump\nobreak}
\def\acknowledgments{\goodbreak\null\jump\nobreak
\itemitem{ }{\ttaarr Acknowledgments} \nobreak\jump\sjump\nobreak}
\def\setchap#1{\ifnum\indappcount=0{\ifnum\subcpcount=0%
\xdef\spzzttrra{\the\cpcount}%
\else\xdef\spzzttrra{\the\cpcount .\the\subcpcount}\fi}
\else{\ifnum\subcpcount=0 \xdef\spzzttrra{\applett}%
\else\xdef\spzzttrra{\applett .\the\subcpcount}\fi}\fi
\expandafter\xdef\csname#1\endcsname{\spzzttrra}}
\newcount\draftnum  \global\draftnum=0
\catcode`@=11
\gdef\Ref#1{\expandafter\ifx\csname @rrxx@#1\endcsname\relax%
{\global\advance\countref by 1%
\ifnum\countref>200%
\message{>>> ERROR: maximum number of references exceeded <<<}%
\expandafter\xdef\csname @rrxx@#1\endcsname{0}\else%
\expandafter\xdef\csname @rrxx@#1\endcsname{\the\countref}\fi}\fi%
\ifnum\draftnum=0 \csname @rrxx@#1\endcsname \else#1\fi}
\gdef\beginref{\ifnum\draftnum=0  \gdef\Rref{\fairef}
\gdef\endref{\scriviref} \else\relax\fi}
\gdef\Rref#1#2{\item{[#1]}{#2}}  \gdef\endref{\relax}
\newcount\conttemp
\gdef\fairef#1#2{\expandafter\ifx\csname @rrxx@#1\endcsname\relax
{\global\conttemp=0
\message{>>> ERROR: reference [#1] not defined <<<} } \else
{\global\conttemp=\csname @rrxx@#1\endcsname } \fi
\global\advance\conttemp by 50
\global\setbox\conttemp=\hbox{#2} }
\gdef\scriviref{\clearconnttrre\conttemp=50
\loop\ifnum\connttrre<\countref \advance\conttemp by 1
\advance\connttrre by 1
\item{[\the\connttrre]}{\unhcopy\conttemp} \repeat}
\clearcountref \clearconnttrre
\catcode`@=12
\ifx\oldchi\undefined \let\oldchi=\chi
  \def\cchi{{\raise 1pt\hbox{$\oldchi$}}} \let\chi=\cchi \fi
\def\del{\partial}   
\def\frac#1#2{{\textstyle{#1 \over #2}}}

\def\half{\ifinner {\scriptstyle {1 \over 2}}\else {1 \over 2} \fi}
  
\def\vev#1{\left\langle#1\right\rangle}

\def\buildchar#1#2#3{{\null\!\mathop{#1}\limits^{#2}_{#3}\!\null}}

\null
%
%--------------------------------------------------------------------
%
%                        PREPRINT FOLLOWS
%
%--------------------------------------------------------------------
%
\def\ttaarr{\bfone} \def\ppaarr{\tencaps}
\def\pb{\overline\psi} \def\ttau{\tilde\tau}
\ifnum\ingrandimento=1095 \font\frxft=cmsy10 at 15.774pt \else
\font\frxft=cmsy10 at 17.28pt \fi
\def\nefreccia{\hbox{\frxft\char'45}}
\def\sefreccia{\hbox{\frxft\char'46}}
\nopagenumbers
\line{\hfill IASSNS-HEP-91/59}
\line{\hfill PUPT-1280}
\line{\hfill September, 1991}
\ifnum\unoduecol=2 \bjump\bjump\else\vfill\fi
\centerline{\capsone HERMITIAN vs. ANTI-HERMITIAN 1-MATRIX}
\sjump
\centerline{\capsone MODELS AND THEIR HIERARCHIES}
\bjump\bjump
\centerline{\tencaps Timothy Hollowood\footnote{$^*$}{Address after
Oct. 1, 1991: Dept. of Theoretical Physics, Oxford, U.K.}, Luis
Miramontes\footnote{$^\dagger$}{Address after Oct. 1, 1991: CERN,
Geneva, Switzerland.} \& Andrea Pasquinucci}
\sjump
\centerline{\sl Joseph Henry Laboratories, Department of Physics,}
\centerline{\sl Princeton University, Princeton, N.J. 08544}
\bjump
\centerline{\tencaps Chiara Nappi}
\sjump
\centerline{\sl Institute for Advanced Study,}
\centerline{\sl Olden Lane, Princeton, N.J. 08540.}
\vfill
\ifnum\unoduecol=2 \eject\null\vfill\fi
\centerline{\capsone ABSTRACT}
\sjump
\noindent Building on a recent work of \v C.~Crnkovi\'c,
M.~Douglas and G.~Moore, a study of
multi-critical multi-cut one-matrix models and their associated
$sl(2,{\mmmath C})$ integrable hierarchies, is further pursued.
The double scaling limits of hermitian matrix models with different
scaling ans\"atze, lead, to the KdV hierarchy, to the
modified KdV hierarchy and part of the
non-linear Schr\"odinger hierarchy. Instead, the anti-hermitian
matrix model, in the
two-arc sector, results in the Zakharov-Shabat hierarchy, which
contains both KdV and mKdV as reductions.
For all the hierarchies, it is found that the Virasoro constraints
act on the associated tau-functions.
Whereas it is known that the ZS and KdV models lead to
the Virasoro constraints of an $sl(2,{\mmmath C})$ vacuum, we find
that the mKdV model leads to the Virasoro constraints of a
highest weight state with arbitrary conformal dimension.
\sjump
\ifnum\unoduecol=2 \vfill\fi
\eject
\yespagenumbers\pageno=1
\introduction
Most of the
interesting properties of the matrix model formulation of two
dimensional gravity were originally extracted for the special case of
lagrangians with even potentials [\Ref{oldmat}].
In the hermitian 1-matrix model it was later shown that the
introduction of odd terms in the potential
gives rise to a doubling of the critical degrees of freedom and a
doubling of the critical equations. In the 1-arc sector one gets two
decoupled Painlev\'e I equations, for the first critical point; the
underlying integrable structure being
two decoupled Korteweg-de Vries (KdV) hierarchies
[\Ref{witten},\Ref{petropoulos}]. The situation, however,
turns out to be different in the 2-arc sector of the theory,
which, for an even potential, has a
Painlev\'e II equation as the lowest multi-critical point; the
underlying integrable structure being the
{\it modified\/} Korteweg-de Vries (mKdV)
hierarchy [\Ref{mkdvmm},\Ref{twoarc}]. Explicit
calculations show that the introduction of odd terms do not lead to
decoupled equations and to the doubling of the mKdV system
[\Ref{nappi}].

This paper originated from our attempt to explore the integrable
structures associated with the 2-arc sector of the hermitian
1-matrix model, with a generic potential. Indeed, one of the most
interesting features of matrix models is the fact that the known
2d quantum gravity models (both pure and
coupled to minimal conformal matter) are described by an integrable
hierarchy, supplemented with an additional condition known as the
`string equation' [\Ref{douglas}].
The common belief is that the
(anti-) hermitian $n$-matrix model should correspond to a hierarchy
associated to the Lie algebra $sl(n+1,
{\mmmath C})$, in the sense that, choosing different scaling
ans\"atze for the double scaling limit of the matrix model, one gets
a field theory, where the free energy (or a function related to it),
satisfies the string equation of such a hierarchy.
Obviously, the case $n=1$ is the simplest and
also virtually the only one where examples can be worked out
explicitly. More general one-matrix models have also been considered
in ref. [\Ref{morris}], which deals with the case for complex
matrices.

We start by deriving in an explicit way the higher multi-critical
points of the hermitian 1-matrix model in the 2-arc sector
with generic polynomials. We find
that the resulting string equations are associated with only the
`even' subset of
flows of the non-linear Schr\"odinger (NLS) hierarchy. This, indeed,
seems to be the hierarchy behind the 2-arc sector of the
hermitian 1-matrix model, as one can check by computing explicitly
the Lax operator.

Interestingly, we notice that if the odd terms in the potential were
purely imaginary, i.e. if the starting matrix model were
anti-hermitian instead of hermitian, the multi-critical points
in the 2-arc sector reproduce all the string equations associated to
the ZS hierarchy. Indeed the Lax operator one gets in this case is
that of the Zakharov-Shabat (ZS) hierarchy\note{The connexion of the
ZS hierarchy with the (anti-) hermitian matrix model in the 2-arc
sector was first noticed by the authors of ref. [\Ref{moore}].}.
The reason why we do not find the odd string equations of the NLS
in our derivation of multi-critical points of the hermitian
matrix model is that those equations are complex. Instead all the
flows of the ZS hierarchy are real as we will see in section 2.5.

While the NLS hierarchy is known to contain, as reduction, the mKdV
hierarchy,
the ZS hierarchy  contains both the mKdV and KdV
hierarchies, [\Ref{hier}], (which can equivalently be obtained
directly from the matrix model by restricting the ansatz
made for the double scaling limit). In addition, we find that the
string equations of the NLS hierarchy reduce to those of the mKdV
hierarchy, and the string equations of ZS to those of the KdV
and mKdV hierarchies.
{}From this one deduces that solutions of the KdV and mKdV
theories correspond also
to solutions of the NLS and ZS theories, in a particular subspace
spanned by the even flows of the two hierarchies. The results are
conveniently illustrated in diagram 1, which shows how the critical
points of the various hierarchies relate both in the 1-arc and
2-arc sectors.

\midinsert
\hfuzz=60pt
$$
\matrix{   &        &  {\rm KdV} &               &         &\cr
&\nefreccia\hbox to 20pt{\smash{{\sevenrm\raise-.5ex\hbox{even
p.}}}}&&&&\cr
\hbox{\rm 1-arc}&   &            &               &         &\cr
&\sefreccia\hbox to 20pt{\smash{{\sevenrm\raise1.0ex\hbox{general
p.}}}}&&&&\cr
&&\buildchar{\hbox{\rm KdV}}{ }{\hbox{\sevenrm (double)}}&&&\cr
           &        &            &               &         &\cr
           &        &            &               &         &\cr
           &        &            &               &         &\cr
           &        &            &               &         &\cr
&&&\buildchar{\hbox{\rm NLS}}{ }{\hbox{\sevenrm (even)}}&
 \buildchar{\longrightarrow}{ }{\hbox{\sevenrm red.}}&
  {\rm mKdV}\cr
&&\nefreccia\hbox to 24pt{\smash{{\sevenrm\raise-.5ex\hbox{general
 p.}}}}&&&\cr
&&&&&\cr
&\nefreccia\hbox to 20pt{\smash{{\sevenrm\raise -.5ex\hbox{herm.}}}}&
 \sefreccia\hbox to 24pt{\smash{{\sevenrm\raise1.0ex\hbox{even
  p.}}}}&&&\cr
&&&&&\cr
\hbox{\rm 2-arc}&   &            & {\rm mKdV}         & &   \cr
&&&&&\cr
&\sefreccia\hbox to 20pt{\smash{{\sevenrm\raise1.0ex\hbox{anti-h.}}}}&
 \nefreccia\hbox to 24pt{\smash{{\sevenrm\raise-.5ex\hbox{even
  p.}}}}&&&\cr
&&&&&\cr
&&\sefreccia\hbox to 24pt{\smash{{\sevenrm\raise1.0ex\hbox{general
  p.}}}}&&
 &{\rm KdV}\cr
        &   &        &            &\nearrow       &         \cr
        &   &        &{\rm ZS} &\hfill\hbox{\sevenrm red.}& \cr
        &   &        &            &\searrow       &         \cr
        &   &        &            &              &{\rm mKdV}\cr}
$$
\centerline{Diagram 1}
\hfuzz=0.1pt
\endinsert

We then go on to discuss the tau-function formalism of the ZS
hierarchy. In this case we find that the partition function of the
theory is equal to the tau-function of the hierarchy, rather than to
its square, as happens instead in the 1-arc KdV model.
As it was first shown in ref. [\Ref{moore}], the partition
function of the matrix model leading to the ZS
hierarchy, satisfies the Virasoro constraints
of an untwisted boson with an, a priori, arbitrary value for
the zero-mode or
`momentum'. This implies that the Virasoro constraints
act on the tau-function, to mirror the situation for the 1-arc KdV
model [\Ref{vircostr}].
However, in the ZS case, the tau-function carries an additional
quantum number due to the zero-mode of the untwisted field.
This additional quantum number seems to play the r\^ole of a
non-perturbative parameter which labels different sectors of the
theory and arguments connected with the tau-function formalism suggest
that it takes discrete values.

An interesting side issue concerns the existence of Virasoro
constraints for the models described by the mKdV hierarchy; these
include both the 2-arc (anti-) hermitian model, with even potential,
and the unitary matrix models [\Ref{unimoore},\Ref{periwal}].
We find that there are, indeed, Virasoro constraints for the
mKdV model; however, the situation is more complicated than in the KdV
case. In the KdV case, the Virasoro constraints act on the tau-function:
that is the square root of the partition function. For the mKdV model
there are two tau-functions; the partition function being the product.
We find that the string equation and the mKdV hierarchy imply a set of
Virasoro constraints for each tau-function separately, however, in
contrast to the KdV case, only the $L_n$ for $n\geq0$ appear, and
the eigenvalue of $L_0$ is an, {\it a priori\/}, undetermined
integration constant. In other words, the tau-functions satisfy the
Virasoro constraints of a highest weight state of the conformal
algebra.

Under the reduction from ZS to KdV, the Virasoro constraints are
transformed into those of a twisted boson acting on the square root of
the partition function (as expected [\Ref{vircostr}]).
More interestingly, the reduction to KdV of the $t_2=\ $constant,
$t_0=x$, ZS scaling theory gives rise to the topological point of the
KdV hierarchy! In other words, from the anti-hermitian 1-matrix model
with a potential of the fourth order it is possible, after the double
scaling limit, to get Topological Gravity [\Ref{topwitten}].

For the reduction to mKdV we get a series of constraints acting on the
mKdV partition function which are not
Virasoro-like; this is to be expected because they act on the product
of the mKdV tau-functions and not on each of them separately. We also
show
that the `conformal dimension' of the mKdV partition function is
fixed under the reduction. Finally in both reductions the value of
the above-mentioned discrete parameter is fixed.

Another important issue which arises from this analysis, is the
possible
existence of new continuum theories described by the `odd' critical
points of the ZS model. The first such model was discussed in
[\Ref{moore}], and it seems to display a topological nature, similar
to the topological point in the 1-arc KdV model [\Ref{topwitten}].
Nevertheless, the theory is sufficiently different from the `normal'
topological theory, to make it unclear as to what continuum theory it
describes. However, like its 1-arc cousin, this topological
theory cannot be obtained from the matrix model.
The possible higher `odd' multi-critical points have yet to
be explored.
We will not address the mathematical technicalities required to prove
that solutions of these new scaling points exist, however, since the
mathematical apparatus exists
[\Ref{bigmoore}] we hope these questions will be tackled elsewhere.

Finally we notice that models with anti-hermitian
matrices seem to arise in relation with topological gravity
(the Kontsevich model [\Ref{konts}]) and with the Penner model
[\Ref{penner}].
Indeed, whereas hermitian matrix models naturally emerge from the
study of quantum gravity as a theory of random surfaces [\Ref{oldmat}],
anti-hermitian matrices seem to arise when one tries
to make connexions between matrix models and moduli spaces of Riemann
surfaces.
\chapter{(Anti-) Hermitian 1-Matrix Models and Orthogonal Polynomials}
In this section we will consider the double scaling limit of a general
(anti-) hermitian 1-matrix model. Most of what follows has been already
extensively discussed in the literature, so we will recall only the
most important results, add some new ones and develop a few explicit
examples to fix our notation.
\section{Double Scaling Limit and String Equations of Hermitian
Matrix Models}
Let $M$ be a hermitian $N\times N$ matrix and consider
$$
{\cal Z}_N\ = \ \int dM e^{-\beta {\rm tr}V(M)}
\efr
where $V(\lambda)=g_1 \lambda + {g_2\over 2}\lambda^2 + {g_3\over 3}
\lambda^3 + \dots $ and $\lambda$ denotes a (real) eigenvalue of $M$.

As it is well known, one can introduce orthonormal polynomials
${\cal P}_n (\lambda)$ such that
$$
\int d\lambda e^{-\beta V(\lambda)} {\cal P}_n (\lambda) {\cal P}_m
(\lambda)\  = \ \delta_{n,m}
\efr
with
$$
\lambda{\cal P}_n(\lambda)\ =\ \sqrt{R_{n+1}}\,{\cal P}_{n+1}(\lambda)
+ S_n {\cal P}_n(\lambda) + \sqrt{R_n}\, {\cal
P}_{n-1}(\lambda)\quad,
\nfr{unop}
then
$$
{\cal Z}_N\ =\ N!\ \prod_{i=0}^{N-1}h_i \ =\ N! \ h_0^N\
\prod_{i=1}^{N-1} R^{N-i}_i
\nfr{unoc}
where $ R_n = h_n / h_{n-1}$.

We will consider, for the moment, the case with real potentials.
Notice that this implies that both $R_n$ and $S_n$ are
real. Moreover, if $V(\lambda)=V(-\lambda)$ then $S_n = 0$.

The most important equation is the so-called `string equation', which
can be written as:
$$
\eqalign{
{n\over \beta}\  &=\  g_2 R_n + g_3(R_nS_n+R_n S_{n-1}) + \cdots \cr
0\  &=\  g_1 + g_2 S_n + g_3 (S^2_n + R_n + R_{n+1}) + \cdots
\ .\cr}
\nfr{unoa}
In the double scaling limit one assumes that $\beta/N\rightarrow 1$,
${n / \beta}  =  1 - {x / N^\alpha}$ and
$$
\eqalign{
R_n\  &=\ 1 + (-1)^n {f(x)\over N^{\gamma_1}} + {r(x)\over
N^{\gamma_2}} + {j(x)\over N^{\gamma_3}} + {z(x)\over N^{\gamma_4}}
+ \cdots \cr
S_n \ &=\ b + (-1)^n {g(x)\over N^{\delta_1}} + {s(x)\over
N^{\delta_2}} + {p(x)\over N^{\delta_3}} + {v(x)\over N^{\delta_4}}
+ \cdots \cr}
\nfr{unob}
where $b$ is an arbitrary constant.

For an even potential, $V(\lambda)=V(-\lambda)$, and $V(\lambda)={g_2
\over 2} \lambda^2 + \cdots + {g_{2k}\over 2k} \lambda^{2k}$, one sets
in the 1-arc sector
$$
f(x)=0=g(x)\ , \qquad \alpha={2k\over 2k+1}\ ,\qquad\gamma_i=\delta_i=
{i\over 2k+1}
\efr
and in the 2-arc sector
$$
\alpha={2k-2\over 2k-1}\ ,\quad\qquad
\gamma_i=\delta_i={i\over 2k-1}\ ,\efr
and one needs to introduce scaling functions up to the order $2k-2$.

Let us introduce $a$ such that\note{This is not the
`a' defined by $Na^{2+1/m}=1$ usually introduced in the literature.}
$$
\eqalign{ a\ &=\ N^{1\over 2k+1} \qquad\quad \hbox{\rm in the 1-arc
sector}\cr
a\ &=\ N^{1\over 2k-1} \qquad\quad \hbox{\rm in the 2-arc
sector,}\cr}
\efr
then eqs. \unoa\ become
$$
\eqalign{
1-{x\over a^\epsilon} \ &=\ {\cal F}_0(g_i) + {f(x)\over a}
{\cal F}_1(g_i) + {1\over a^2} {\cal F}_2(f,g,r,s,f^\prime,g^\prime
;g_i) + \cdots \cr
0\ &=\ {\cal G}_0(g_i) + {g(x)\over a} {\cal G}_1(g_i) + {1\over a^2}
{\cal G}_2(f,g,r,s,f^\prime,g^\prime ;g_i) +\cdots \cr}
\nfr{unod}
where $\epsilon=2k$ in the 1-arc sector, $\epsilon=2k-2$ in the 2-arc
sector and $\phantom{f}^\prime$ means ${\del\phantom{x}\over \del x}$.
The functions ${\cal F}_i$ and ${\cal G}_i$ can be explicitly obtained
from eqs. \unoa\ and depend on the coupling constants and on the
scaling  functions.

It is now easy to solve order by order in $1/a$ the string equation
\unod, for example the zeroth order in $1/a$ (${\cal F}_0=0={\cal
G}_0$) fixes the value of $g_1$ and
$g_2$\note{Most of the
computations of this section have been done with the help of
{\it Mathematica}$^{\rm TM}$.}:
$$
\eqalign{g_2\ =&\ 1 - 2bg_3 - (3 + 3b^2) g_4 - (12 b + 4 b^3) g_5
- (10 + 30 b^2 + 5 b^4) g_6 + \cdots \cr
g_1\ =&\ -b g_2 - (2 + b^2) g_3 -(6b+b^3)g_4 - (6 + 12 b^2 +b^4)
g_5 + \cr
&\  -(30 b + 20 b^3 + b^5 ) g_6 +\cdots \quad .\cr}
\efr
Notice that the $g_i$'s depend on the free parameter $b$.

At order $1/a$ one finds two
possible solutions: either $f(x)=0$ and $g(x)=0$ or one fixes the
value of $g_4$. Choosing the solution $f(x)=0=g(x)$ one gets the
1-arc sector string equations; fixing $g_4$ instead, one gets the
2-arc string equations. In the case of an even potential of order 4
($k=2$) these are the well-known Painlev\'e I and Painlev\'e II string
equations.

Moreover, from eq. \unoc, in the double scaling limit, we find the
`specific heat' (up to non-universal terms)
$$
F^{\prime\prime}\ =\ \frac{1}{2} f^2(x) - r(x)
\efr
where ${\cal Z}=\hbox{\rm exp}(-F)$ and $\langle PP\rangle =
{\del^2_x}\ {\rm log}\,{\cal Z}=-F^{\prime\prime}$.

Let us consider now a real potential $V(\lambda)$ with both even and
odd couplings. In the 1-arc sector ($f(x)=0=g(x)$) one gets the well
known `doubling' phenomenon. For example, from
$$
V(\lambda)\ = \ (\frac13 b^3-2b)\lambda + \frac12(2-b^2)\lambda^2
+ \frac13 b \lambda^3 - \frac1{12} \lambda^4
\efr
one has the string equations
$$
x\ =\ \frac13 \chi^{\prime\prime} + \chi^2\ ,\qquad\quad
x\ =\ \frac13 \overline\chi^{\,\prime\prime} + \overline\chi^{\, 2}
\efr
where $\chi(x)=r(x)-s(x)$ and $\overline\chi(x)=r(x)+s(x)$. Notice
that these equations do not depend on the parameter $b$ and that the
$g_{2i+1}$ are proportional to $b$. In the same way, in the 1-arc
sector one obtains all the string equations of the KdV hierarchy,
which are described in the following chapter.
The `specific heat' for these models is given by
$F^{\prime\prime} = - r(x) = -{1\over 2} (\chi
+\overline\chi)$.

The 2-arc sector with a general real potential is more interesting,
indeed many different scaling solutions and string equations appear.
For example, with
$$
V(\lambda)\ =\ (2b-b^3)\lambda + \frac12(3b^2 -2)\lambda^2 -
b\lambda^3 + \frac14\lambda^4
\efr
one gets
$$
\eqalign{&f^{\prime\prime} - \frac14 f (g^2 + f^2) + \frac12 f x
\ =\ 0\cr
&g^{\prime\prime} - \frac14 g(g^2+f^2) + \frac12 g x \ =\ 0\cr}
\nfr{unoe}
and from
$$
\eqalign{V(\lambda) \ =\ &(b+b^3 - \frac12 b^5)\lambda +
\frac12 (-1-3b^2 +
\frac52 b^4) \lambda^2 + \frac13 ( 3b - 5b^3)\lambda^3 +\cr
& \frac14 (-1+5b^2)\lambda^4 -\frac12 b\lambda^5 + \frac1{12}
\lambda^6 \cr}
\efr
one gets
$$
\eqalign{0\ =\ &xf - \frac38 f^3(g^2+f^2) - \frac38 fg^2(f^2+g^2)
+ \frac52 f (f^\prime)^2 + 3 g f^\prime g^\prime +\cr
& - \frac12 f (g^\prime)^2 +
\frac52 f^2 f^{\prime\prime} + \frac32 g^2 f^{\prime\prime} + f
g g^{\prime\prime} - f^{(4)}\cr
0\ =\ &xg - \frac38 g^3(g^2+f^2) - \frac38 gf^2(f^2+g^2)
+ \frac52 g (g^\prime)^2 + 3 f f^\prime g^\prime +\cr
& - \frac12 g (f^\prime)^2 +
\frac52 g^2 g^{\prime\prime} + \frac32 f^2 g^{\prime\prime} + f
g f^{\prime\prime} - g^{(4)}\ .\cr}
\nfr{unof}
Again these equations do not depend on the parameter $b$; the
$g_{2i+1}$ are proportional to $b$ and putting $g=0$ these equations
will turn out to be the first two string equations of the
mKdV hierarchy [\Ref{twoarc}]. When $g$ is not put
to zero, these two string equations will be seen, in the next chapter,
to be the {\it second\/} and {\it fourth\/}
string equation of the Non-Linear-Schr\"odinger (NLS) hierarchy.
It is not possible to get the first and third string equation of the
NLS hierarchy from the hermitian 1-matrix model. It seems that the NLS
hierarchy does not have a
full `matrix model' realization, in the sense that we do not
get the multi-critical points
corresponding to the `odd' flows; a fact that will be
explained in section \S2.5.

For all these models the `specific heat' turns out to be
$$
F^{\prime\prime} \ =\ {1\over 4}\left[ f^2(x) + g^2(x) \right]
\efr
since $r(x) = {1\over 4}( f^2(x) - g^2(x))$.
\section{Anti-Hermitian Matrix Models and Their String Equations}
Let us first consider, in more detail, the NLS string equations
\unoe\ and
\unof. Sending $g \rightarrow ig$ one obtains string equations that
are nothing but the string equations associated to the
Zakharov-Shabat (ZS) hierarchy, as will be
apparent from the next chapter. But this substitution is incompatible
with
our ansatz, eq. \unob, for the double scaling limit.
Indeed, sending $g \rightarrow
ig$ implies that $S_n$ becomes complex. But if the potential is real,
$V^*(\lambda)=V(\lambda)$, it is easy to show that $S_n$ must be real.
On the other hand, if $V^*(\lambda)=V(-\lambda)$ then $S_n$ is pure
imaginary. (This can be easily shown using the orthogonality of the
polynomials $P_n(\lambda)$ ($P_n(\lambda)=\sqrt{h_n}
{\cal P}_n(\lambda)$), the fact that $P_n(\lambda)=\lambda^n +
O(\lambda^{n-1})$ and the reality of the partition function
(see \unoc).)
Therefore, we are led to consider the potential
$V(\lambda) = ig_1\lambda + \frac12 g_2 \lambda^2 + \frac13 i g_3
\lambda^3 + \frac14 g_4\lambda^4 + \cdots $
with $\lambda,\, g_k \in {\mmmath R}$ and some of the $g_{2i+1}$
different from zero. The double scaling limit is now done in exactly
the same way as in the previous section except for the ansatz for
$S_n$ (eq. \unob) which is instead pure imaginary, i.e.
$$
S_n\ =\ ib+(-1)^n {ig(x)\over a} + {is(x)\over a^2} +{ip(x)\over a^3}+
{iv(x)\over a^4} + \cdots
\nfr{unos}
with $b,\ g(x), \ s(x), \ \dots$ all real.

Notice that we can get rid of the `$i$' in the potential introducing
{\it anti-hermitian\/} matrices $\widetilde{M}=iM$ with pure imaginary
eigenvalues $\widetilde\lambda=i\lambda$.  Again the $R_n$ are real
and in the same way it is easy to prove that the $S_n$ are pure
imaginary.

For notational simplicity in the rest of this section we will
continue to use hermitian matrices with complex potentials instead of
using directly anti-hermitian matrices.

In the 1-arc sector one gets again the double KdV string equation,
now however in the variables $\chi = r-is$ and
$\overline\chi = r+is=\chi^*$.

In the 2-arc sector we found the family of
scaling solutions corresponding to the ZS hierarchy. For example,
from
$$
V(\lambda)\ =\ i(2b+b^3)\lambda + \frac12 (-2 -3b^2)\lambda^2 -
ib \lambda^3 + \frac14 \lambda^4
\nfr{unozz}
with $\epsilon=2$ one gets
$$
\eqalign{0\ &=\ \frac12 xf + f^{\prime\prime} + \frac12 f ( g^2
- f^2 ) \cr
0\ &=\ \frac12 xg + g^{\prime\prime} + \frac12 g ( g^2
- f^2 )\ . \cr}
\nfr{unoh}
{}From
$$
\eqalign{V(\lambda) \ =\ &i\left[-1 +b + b^2 - b^3 + \frac12 b^4 -
\frac12  b^5\right]\lambda + \cr
& \frac12 \left[ -1 -2b + 3b^2 -2b^3
+ \frac52 b^4\right] \lambda^2 +\cr
& \frac13 i\left[-1 +3b -3b^2 + 5 b^3 \right] \lambda^3 + \frac14
\left[-1+2b -5b^2\right]\lambda^4 +\cr
& \frac15 i \left[ \frac12
-  \frac52 b\right] \lambda^5 + \frac1{12} \lambda^6 \cr}
\nfr{unon}
and $\epsilon=3$ one has
$$
\eqalign{ 0\ &=\ xf + g^{\prime\prime\prime} + \frac32 g^\prime
\left( g^2 -f^2\right) \cr
0\ &=\ xg + f^{\prime\prime\prime} + \frac32 f^\prime
\left( g^2 -f^2\right)\ . \cr}
\nfr{unog}
Finally, from
$$
\eqalign{V(\lambda) \ =\ &i\left[ b - b^3  -
\frac12 b^5\right]\lambda + \frac12 \left[ -1  + 3b^2
+ \frac52 b^4\right] \lambda^2 +
\frac13 i \left[ 3b + 5 b^3 \right] \lambda^3 +\cr
& \frac14
\left[-1-5b^2\right]\lambda^4 - \frac12 i
b \lambda^5 + \frac1{12} \lambda^6 \cr}
\nfr{unoi}
with $\epsilon=4$ one has
$$
\eqalign{0\ =\ & xf + \frac38 f^3 \left( g^2 - f^2 \right) -
\frac38 f g^2 \left( g^2 - f^2 \right) + \frac52 f (f^\prime)^2
- 3g f^\prime g^\prime +\cr
& \frac12 f (g^\prime)^2 +
\frac52 f^2 f^{\prime\prime} - \frac32 g^2 f^{\prime\prime}
-fgg^{\prime\prime} - f^{(4)}\cr
0\ =\ & xg - \frac38 g^3 \left( g^2 - f^2 \right) +
\frac38 g f^2 \left( g^2 - f^2 \right) - \frac52 g (g^\prime)^2
+ 3f f^\prime g^\prime +\cr
& - \frac12 g (f^\prime)^2
- \frac52 g^2 g^{\prime\prime} + \frac32 f^2 g^{\prime\prime}
+fgf^{\prime\prime} - g^{(4)}\ .\cr}
\nfr{unol}
These are the first three string equations of the ZS hierarchy
(excluding the topological-like point [\Ref{moore}] which cannot be
obtained
from the matrix model). More precisely, they correspond to the points
$t_\epsilon=\ $constant, $t_0=x$ and $t_i=0$ for
$i\neq \{0,\epsilon\}$ in the ZS hierarchy (see next section).

A few comments are in order. As usual the string equations do not
depend on $b$. The $g_{2i+1}$ are proportional to $b$ except for
the case of eq. \unon. Indeed, it is not possible to get the
string equation eq. \unog\ from a real potential
($V^*(\lambda)=V(\lambda)$).  Moreover, setting $g=0$ one
obtains the first two string equations of the mKdV hierarchy (eq.
\unog\ vanishes identically) and setting ${1\over 2}(f+g)=-1$ in
eqs. \unoh\ and \unol\ one gets
the first two string equations (topological point included) of the
KdV hierarchy in the variable $\psi= {1\over 2}(f-g)$.

For all these models it turns out that
\fr
F^{\prime\prime}\ =\ -{1\over 4}\left[ g^2(x) - f^2(x) \right]
\nfr{unom}
since $r(x)= {1\over 4} [ f^2(x) + g^2(x)]$.
\section{Lax Operators from Matrix Models}
\setchap{laxop}
The basic idea, due to Douglas [\Ref{douglas}], is that it is possible
to make the double scaling limit
not only on the string equations but also directly on $\lambda$ and
${\del\phantom{\lambda}\over \del\lambda}$ seen as operators acting on
${\cal P}_n(\lambda)$. This operator is then related to the Lax
operator of the integrable
hierarchy which underlies the behaviour of the continuum theory
associated to the particular sequence of scaling ans\"atze
chosen. Indeed, under a double scaling limit from eq.
\unop\ it is easy to see that $\lambda$ becomes a differential operator
($\widehat\lambda$) of order 2 in $x$. The obvious relation
$$
\left[{\del\phantom{\lambda}\over\del\lambda},\lambda\right]\ =\ 1
\efr
after the double scaling limit becomes the string equation
[\Ref{douglas}].
Eq. \unop\ under the double scaling limit becomes
$$
\widehat\lambda \Psi \ =\ \Lambda \Psi
\nfr{unoq}
where $\Psi$, which can be a vector, is related to the rescaled
polynomials and $\Lambda$ is a differential operator
of degree 2 in $x$. Analogously, there exists an equation of the form
$$
{\widehat{\del\phantom\lambda} \over \del\lambda} \Psi \ =\  M
\Psi\quad .
\efr
These two equations can be rewritten as
$$
L\Psi\  \buildchar{=}{\rm def}{ }
\ \left( \widehat\lambda -\Lambda \right)
\Psi \ =\ 0\ ,  \qquad
 \left({\widehat{\del\phantom\lambda} \over \del\lambda} - M \right)
\Psi \ =\ 0
\efr
and then the string equation becomes the compatibility
condition for these differential equations [\Ref{bigmoore}], i.e.
$$
\left[ L, \left({\widehat{\del\phantom\lambda} \over \del\lambda}
- M \right) \right] \ =\ 0\quad .
\efr
It is easy to get the explicit expression of $L$ from eq. \unoq. $L$
is then the Lax operator of the corresponding
hierarchy. Using the techniques of Zakharov-Shabat and
Drinfeld-Sokolov
[\Ref{drinfel},\Ref{zaksab}], one can also explicitly compute $M$
[\Ref{douglas},\Ref{mkdvmm},\Ref{moore}].

We now explicitly compute $L$ from eq. \unop. Consider first the
case of a hermitian matrix with a real potential. Let
$\Pi(x,\lambda)$ denote ${\cal P}_n(\lambda)$ after the double scaling
limit\note{Actually, as explained in refs.
[\Ref{douglas},\Ref{mkdvmm},\Ref{bigmoore}], under the double
scaling limit
$\Pi(x,\lambda) \sim e^{-NV(\lambda)/2} {\cal P}_n(\lambda)$
has a smooth behaviour and should be considered for this
computation.},
thus eq. \unop\ becomes
$$
\lambda\, {\cal P}_n(\lambda)\ \sim\
\ (2+b) \Pi(x) + {1\over a^2} \left[ r(x) \Pi(x) + s(x) \Pi (x) +
\Pi^{\prime\prime} (x) \right] + \cdots \ ,
\efr
where for notational simplicity we have dropped the dependence on
$\lambda$ in $\Pi$. Setting $\lambda -(b+2) \rightarrow
\widehat\lambda/a^2$ one has
$$
L\, \Pi(x,\lambda)\ =\ 0\ ,
\efr
where
$$
L\ =\ {\del^2_x} + \left( r(x) + s(x) \right)
- \widehat\lambda\ .
\efr
This is the Lax operator of the KdV hierarchy.

Notice that setting $\Pi(x,\lambda)  \sim (-1)^n {\cal P}_n(\lambda)$
one gets $L = {\del^2_x} + \left( r(x) - s(x) \right)
+ \widehat\lambda$. Thus there are two KdV Lax operators associated to
the hermitian 1-matrix model in the 1-arc sector, this is nothing but
the doubling phenomenon [\Ref{witten}].

In a similarly way, for the anti-hermitian 1-matrix model in the
1-arc  sector one gets the previous two Lax operators with
$s(x)\rightarrow i s(x)$.

Consider now the case of the 2-arc sector with a hermitian matrix and
a real potential. In order to get a non trivial Lax operator in the
double scaling limit, following [\Ref{mkdvmm}], we need to introduce
two scaling functions depending on parity,
$\Pi(\lambda,x)\sim (-1)^m {\cal P}_{2m} (\lambda) $  and
$\Omega(\lambda,x-{1\over a}) \sim (-1)^m {\cal P}_{2m+1} (\lambda)$
where $x=x_{2m}$. Thus for $n$ even eq. \unop\ becomes
$$
\lambda\, (-1)^{n/2} {\cal P}_n(\lambda) \
\sim\
b \Pi(x) + {1\over a} \left[ g(x) \Pi(x) - f(x) \Omega (x) - 2
\Omega^\prime (x) \right] + \cdots
\efr
and for $n$ odd one has
$$
\lambda\, (-1)^{(n-1)/2}  {\cal P}_n(\lambda) \
\sim\ b \Omega(x) + {1\over a} \left[- g(x) \Omega(x) - f(x)\Pi (x)+2
\Pi^\prime (x) \right] + \cdots\ .
\efr
Thus, setting
$\lambda -b \rightarrow \widehat\lambda /a$ one has
$$
L\,\Psi\ =\ 0
\nfr{unor}
where
$$
L\ =\ {\widehat\lambda\over 2} -{g(x)\over 2}
\left(\matrix{1&0\cr0&-1\cr}\right) + {f(x)\over 2} \left(
\matrix{0&1\cr 1&0\cr}\right) + {\del\phantom{x}\over \del x} \left(
\matrix{ 0&1\cr -1&0\cr}\right)
\efr
and $\Psi=\left({\Pi \atop \Omega}\right)$. Now, after having
conjugated $L$ and rotated $\Psi$, we get eq. \unor\  with
$$
L\ =\ i\sigma_3 \widehat\lambda + \sigma_2 g - \sigma_1 f +
{\del_x}
\efr
and $[\sigma_l,\sigma_j] = i\epsilon_{ljk}\sigma_k$.
As we expected, this is the Lax operator of
the NLS hierarchy.

In the case of an anti-hermitian matrix with pure imaginary
eigenvalues
$\widetilde\lambda$ one can do the same computation with
${\cal P}_{2m+1}$ and $S_n$ pure imaginary (see eq. \unos). Letting
$\widetilde\lambda -ib \rightarrow  i \widehat\lambda /a$, one finally
gets
$$
L\ =\ {\del_x}
 + i\sigma_2 g - \sigma_1 f - \sigma_3 \widehat\lambda\ ,
\efr
and this is the Lax operator of the ZS hierarchy.

Notice that the mKdV string equations can be obtained both from the
NLS (hermitian matrix, real potential) and ZS (anti-hermitian matrix,
complex potential) hierarchy. This is obvious since setting
$S_n=0$ ($b=0=g=\dots$) hermitian and anti-hermitian matrix models
coincide.
\chapter{Hierarchies and String Equations}
In this section we will discuss the hierarchies of integrable
equations which lie behind the non-perturbative
structure of the one matrix model.
The relevant hierarchies are well-known in the mathematical
physics literature, and are the Zakharov-Shabat (ZS), for the Lie
algebra $sl(2,{\mmmath C})$, the Korteweg-de Vries (KdV),
{\it modified\/} Korteweg-de Vries (mKdV) and non-linear Schr\"odinger
(NLS) hierarchies [\Ref{hier}].

Interestingly, all these hierarchies are intimately related to the Lie
algebra $sl(2,{\mmmath C})$. This is apparent in the matrix
Lax formalism [\Ref{drinfel}] and also the `Hirota'
or `tau-function' formalism [\Ref{kacwak}]. It is also significant
that the
complex-ZS hierarchy is the `master' hierarchy for $sl(2,{\mmmath
C})$, since
the ZS, NLS, KdV and mKdV hierarchies can all be obtained by
appropriate
reductions. This will be explained in more detail below. We shall also
discuss how the hierarchies are related to the matrix models.
\section{The $sl(2,{\mmmath C})$ Hierarchies}
In section \S\laxop\ we showed how the matrix models are connected
with
the integrable hierarchies presented in the Lax formalism. In fact,
the most economical way of explicitly introducing the hierarchies is
via their recursion relations. These can be extracted from the Lax
formalism, see for example ref. [\Ref{drinfel}]. In what follows we
often write $x$ for the flow $t_0$, of the hierarchy under discussion.
For the KdV equation the hierarchy can be presented, in the
following way
$$
{\partial u\over\partial t_k}=\partial_xR_{k+1}\ \ \ k\geq0,
\nfr{duec}
where $R_k$ is a polynomial in $u$ and its $x$-derivatives (the
$R_k$'s are known as the Gel'fand-Diki polynomials [\Ref{GD}], and are
not to be confused with the $R_k$ used in the previous section). The
integrability of the hierarchy is summed up in the recursion relation
$$
\partial_xR_{k+1}=\left(\frac12\partial^3_x
+2u\partial_x+u'\right)R_k\ .
\nfr{kdvr}
Specifying $R_0=1$ along with the recursion relation
completely determines the hierarchy. To make contact with the notation
used in the previous section, for example for the hermitian matrix
model in the 1-arc sector with even potential, one should set $u=r$.

For the complex-ZS hierarchy a similar structure of recursion
relations is
found. For  two independent complex variables $\psi$ and $\pb$, if
$$
{\partial\psi\over\partial t_k}=\frac12(F_{k+1}-G_{k+1})\ ,
\qquad\quad
{\partial\pb\over\partial t_k}=\frac12(F_{k+1}+G_{k+1})\ ,
\efr
with $k\geq-1$, then
$$
\eqalign{F_{k+1}&=G'_k+(\pb-\psi)H_k\cr
G_{k+1}&=F'_k+(\pb+\psi)H_k\cr H_k'&=\pb(G_k-F_k)-\psi(G_k+F_k)\ .\cr}
\nfr{zsr}
Specifying $F_0=\pb-\psi$ and $G_0=\pb+\psi$ along with the recursion
relations
then determines the whole hierarchy. Notice that the flow $t_{-1}$ is
particularly simple
$$
{\partial\psi\over\partial t_{-1}}=-\psi,\ \ {\partial\pb\over\partial
t_{-1}}=\pb\ ,
\nfr{tmodep}
implying the following dependence on $t_{-1}$:
$\psi\sim e^{-t_{-1}}$ and $\pb\sim e^{t_{-1}}$. To make contact with
the notation of the previous section $\psi=\frac12(f-g)$ and
$\pb=\frac12(f+g)$, which also
agrees with the conventions of ref. [\Ref{moore}] when $t_k\rightarrow
t_{k+1}$. The NLS hierarchy is recovered from the complex-ZS hierarchy
by choosing $\pb$ to be the complex conjugate of $\psi$ (up to a term
involving $t_{-1}$), a choice which is
easily seen to be consistent with the recursion relations of the
hierarchy.
The ZS hierarchy itself,  simply corresponds to the
complex-ZS hierarchy with $\psi$ and $\pb$ both real.

The mKdV hierarchy is best introduced through its relation to the KdV
hierarchy via the {\it Miura Map\/}. This map takes a solution of the
mKdV
$\nu$ hierarchy into a solution of the KdV $u$ hierarchy as
$$
u=-\nu'-\nu^2\ .
\efr
The pull-back of the KdV flow $t^{\rm KdV}_k$ under the Miura Map is
then the mKdV flow $t^{\rm mKdV}_k$. We now make this more explicit.
Defining $D=-\partial_x-2\nu$ and $D^\star=\partial_x-2\nu$, the
recursion relation \kdvr\ becomes $\partial_xR_{k+1}=D(-\frac12
\partial_x)D^\star R_k$ [\Ref{wilson}]. Since
$\partial_xu=D\partial_x\nu$, we have
$$
{\partial u\over\partial t_k^{KdV}}=D{\partial\nu\over\partial
t_k^{mKdV}}=D\left(-\frac{1}{2}\partial_x\right)D^\star R_k\ ,
\efr
and so the mKdV flows are
$$
{\partial\nu\over\partial t_k}=-\frac{1}{2}\partial_xD^\star R_k\ \ \
k\geq0,
\nfr{dued}
where the polynomial $R_k=R_k[-\nu'-\nu^2]$
is expressed in terms of $\nu$ and its
$x$-derivatives by substituting $u$ for $\nu$ via the Miura Map. To
make contact with the notation of the previous section $\nu=f/2$.
\section{The ZS, NLS, KDV and MKDV Hierarchies as Reductions of the
complex-ZS Hierarchy}
\setchap{redsec}
In this section we explain how the the various hierarchies that we
have introduced can all be obtained as reductions of the complex-ZS
hierarchy. The situation is conveniently summarized by diagram 2
which shows how the various hierarchies are related by reduction.
\midinsert
$$
\matrix{    &        &{\rm NLS}&\longrightarrow&{\rm mKdV}\cr
            &        &         &               &          \cr
            &\nefreccia&       &               &          \cr
\hbox{\rm complex-ZS}&  &      &               &          \cr
            &\sefreccia&       &               &{\rm KdV} \cr
            &        &         &\nearrow       &          \cr
            &        &{\rm ZS} &               &          \cr
            &        &         &\searrow       &          \cr
            &        &         &               &{\rm mKdV}\cr}
$$
\centerline{Diagram 2}
\endinsert
We have already noted how the complex-ZS hierarchy is reduced to the
ZS and NLS hierarchies by taking two different `real slices'. For the
former one takes $\psi$ and $\pb$ to be real, whilst for the latter
one takes the complex conjugate of $\psi$ to be
$\psi^\star=e^{-2t_{-1}}\pb$.

The KdV hierarchy is obtained from the ZS hierarchy by setting
$$\pb=-e^{t_{-1}},
\nfr{kdvcond}
the KdV variable being given by $u=-\psi\pb$.
One can readily prove that
$F_{2k+1}+G_{2k+1}=0$, when evaluated at $\pb=-e^{t_{-1}}$, whereas
$F_{2k}+G_{2k}\neq0$. This means that only the even flows preserve the
condition \kdvcond, and so only the even flows reduce to flows of the
KdV hierarchy. One finds
$$
\left.{\del(-\psi\pb)\over\partial
t_{2k}}\right\vert_\star=\frac{1}{2}\left.H_{2k+1}'\right\vert_\star,
\efr
where $\star$ means `evaluate at $\pb=-e^{t_{-1}}$'. If one now
compares the recursion relations of the ZS hierarchy \zsr, evaluated
at $\pb=-e^{t_{-1}}$, to those of the KdV hierarchy \kdvr, then one
deduces that
$$H_{2k+1}[\psi,\pb=-e^{t_{-1}}]=2^{k+1}R_{k+1}[-\psi\pb].
\efr
This means that the relation between the flow variables is
$t^{\rm ZS}_{2k}=2^{-k}t^{\rm KdV}_k$. It is straightforward to see
that the KdV hierarchy cannot be obtained from the NLS hierarchy by a
similar reduction.

The mKdV hierarchy is obtained from both the NLS and ZS hierarchies by
setting
$$\pb=e^{2t_{-1}}\psi,
\nfr{mkdvcond}
the mKdV variable being given by $\nu^2=\psi\pb$.
One finds that $G_{2k+1}=F_{2k}=H_{2k}=0$,
when evaluated at \mkdvcond. So the situation is similar to that for
the KdV reduction, in that only the even flows preserve the reduction.
By pursuing a similar analysis of the recursion relations,
one discovers that the flow variables are related via
$t^{ZS}_{2k}=2^{-k}t^{mKdV}_k$ since $F_{2k+1} = -2^k
\del_x D^\star R_k [-\nu'-\nu^2]$, when evaluated at \mkdvcond.
\section{The String Equations and Their Reductions}
The hierarchies that we have discussed above admit many types of
solution. However, in applications
to matrix models and two-dimensional field theories, very particular
solutions are required. In addition to boundary conditions, these are
specified by adjoining to the
hierarchy an extra condition called the {\it string equation\/}
[\Ref{douglas}]. The string equation must be consistent with the flows
of the
hierarchy, in the sense that it must be preserved by the flows of the
hierarchy. It turns out that the string equation admits certain
scaling, or
multi-critical solutions. It is these solutions which are found in the
matrix model, after the double scaling limit. They are obtained by
restricting to the subspace $t_k=\ $constant, $t_0=x$ and $t_j=0$
otherwise, for some $k$. We call the resulting reduced
equation the $k^{th}$ string equation. The string equations can be
found in general following the analysis of \S\laxop.

The string equation associated to the KdV hierarchy was found
originally in [\Ref{douglas}]. In our conventions
it takes the form
$$
\sum_{k=1}^\infty(2k+1)t_k{\partial u\over\partial t_{k-1}}=-1\ ,
\nfr{kdveq}
which may be integrated using \duec\ to give
$$
\sum_{k=0}^\infty(2k+1)t_kR_k=0\ .
\nfr{intstreq}
The $k^{th}$ multi-critical point corresponding to $t_k=\ $constant,
$t_0=x$, and $t_j=0$ otherwise, is described by the string equation
$$(2k+1)t_kR_k[u]=-x\ .
\efr

For the ZS hierarchy, the string equation was found in [\Ref{moore}]
$$
\sum_{k=0}^\infty(k+1)t_k{\partial\psi\over\partial
t_{k-1}}=0,\qquad \sum_{k=0}^\infty(k+1)t_k{\partial\pb\over\partial
t_{k-1}}=0\ .
\nfr{strineq}
The $k^{th}$ multi-critical point for which $t_k=$constant, $t_0=x$
and $t_j=0$ otherwise, is described by the string equation
$$(k+1)t_k(F_k-G_k)=2x\psi,\qquad\quad (k+1)t_k(F_k+G_k)=2x\pb\ .
\efr
The above equations also apply to the NLS hierarchy by taking
$\psi^\star=e^{-2t_{-1}}\pb$.

The string equation of the mKdV hierarchy has been obtained in
[\Ref{mkdvmm}], however, one can obtain it in a simple way given the
string equation of the KdV hierarchy by using the Miura map. The idea
is to
pull back the string equation of the KdV hierarchy via the Miura map
$u=-\nu'-\nu^2$; the result is then guaranteed to be consistent with
the flows of the mKdV hierarchy, because of the Hamiltonian property
of the Miura map. We act on \intstreq\ with the operator $\frac12
\partial_x D^\star$ and use eq. \dued\ to obtain
$$
\sum_{k=1}^\infty(2k+1)t_k{\partial\nu\over\partial t_k}=-x\nu'-\nu\ .
\efr
This can be rewritten as
$$
\sum_{k=0}^\infty (2k+1)t_k{\partial\nu\over\partial t_k}+\nu=0\ .
\nfr{nustr}
Using
eq. \dued\ again, we can integrate with respect to $x$ (discarding, an
integration constant) obtaining the mKdV string equation
\fr
\sum_{k=0}^\infty (2k+1)t_kD^\star R_k=0.
\nfr{mkdveq}
where $R_k = R_k[-\nu'-\nu^2]$. The $k^{th}$ multi-critical point for
which $t_k=\ $constant, $t_0=x$ and otherwise $t_j=0$, is
described by the string equation
$$(2k+1)t_kD^\star R_k[-\nu'-\nu^2]=2x\nu\ .
\efr

We now show that the string equation of the ZS hierarchy consistently
reduces to the string equations of the KdV and mKdV hierarchies,
respectively, for the reductions we discussed in \S\redsec.
For the reduction to KdV we evaluate
\strineq\ at $\pb=-e^{t_{-1}}$,  $t_{2k+1}=0$
and use the relations obtained in \S\redsec\  to get
\fr
\eqalign{ &\sum_{k=0}^\infty (k+\frac{1}{2}) t_k^{\rm KdV} 2^{-k}
\left( \frac12 \del^2_x + \psi\right) H_{2k-1}\ =\ 0\cr
&\sum_{k=0}^\infty (k+\frac{1}{2}) t_k^{\rm KdV} 2^{-k} H_{2k-1}\
=\ 0\ .\cr}
\efr
These equations are obviously equivalent to the KdV string equations
\kdveq\ since $H_{2k-1}=2^k R_k$ at $\pb=-e^{t_{-1}}$, $t_{2k+1}=0$.

Analogously, in the case of the mKdV reduction with
$\pb=e^{2t_{-1}}\psi$,
$t_{2k+1}=0$, one of the equations \strineq\ is trivially solved and the
other equation becomes
\fr
\sum_{k=0}^\infty (k+\frac{1}{2}) t_k^{\rm mKdV} 2^{-k} G_{2k}\
=\ 0\ .
\efr
But $G_{2k}= -2^k D^* R_k[-\nu'-\nu^2]$ at $\pb=e^{2t_{-1}}\psi$,
$t_{2k+1}=0$, which gives the mKdV string equation \mkdveq.

The significance of the fact that the string equation of the ZS
model
reduces to that of the KdV and mKdV models, is that solutions of the
KdV and mKdV string equations are, by pulling back, solutions
of the ZS string equation (on the subspace $t_{2k+1}=0$ $\forall k$).
This will lead us to conclude that the model
described by the ZS hierarchy includes the models described by the
KdV
and mKdV hierarchies. We further discuss these facts in \S3.4 and
\S3.5.
\section{The Tau-Function Formalism}
\setchap{tausec}
There is an alternative formalism for constructing integrable
hierarchies which was originally developed as a direct solution
technique of the non-linear equations of the hierarchy.
The central objects of this approach are the tau-functions,
which satisfy a hierarchy of non-linear `Hirota bilinear equations',
see [\Ref{kacwak}] for example.
For the hierarchies that we are considering, the Hirota hierarchies
are
intimately related to the Lie algebra $sl(2,{\mmmath C})\ (=A_1)$.
In fact, they use the two
vertex operator representations of the basic representations of the
affine algebra $A_1^{(1)}$. This works in the following way. The basic
representations of $A_1^{(1)}$ are carried by the Fock space of a
scalar field, either twisted or untwisted. The Hirota equations for
the
tau-function are equivalent to the condition that the
tau-function lies in the orbit of the highest weight state of the group
associated to the affine algebra. The untwisted construction underlies
the ZS
and NLS hierarchy, whilst the twisted construction underlies the KdV
and mKdV hierarchies [\Ref{kacwak}].

For the joint KdV and mKdV system (related by the Miura Map) there are
two
tau-functions $\tau_0$ and $\tau_1$ arising from the two basic
representations of the Kac-Moody algebra $A_1^{(1)}$.
The relationships between these and the functions $u$ and $\nu$ are
$$
u=2\partial_x^2\,{\rm log}\,\tau_0\ ,\qquad \nu=\partial_x{\rm log}
\left({\tau_1\over\tau_0}\right)\ .
\nfr{utau}
One of the equations of the hierarchy is
$$
\tau_0''\tau_1-2\tau_0'\tau_1'+\tau_0\tau_1''=0\ ,
\nfr{reltau}
from which one can extract the Miura Map $u=-\nu'-\nu^2$ and the
relation
$$
\nu^2=-\partial_x^2\,{\rm log}\,\left(\tau_0\tau_1\right)\ .
\nfr{tausq}

The Hirota hierarchy which leads to the complex-ZS hierarchies has an
infinite
set of tau-functions. This is because the relevant vertex operator
construction, in this case, involves an untwisted scalar field, which
has a zero-mode. In order that the operator product expansions of the
vertex operators are local, the zero-mode must be quantized, taking
values in the weight lattice of the finite Lie algebra; $A_1$ in this
case, whose weight
lattice is simply isomorphic to $\mmmath Z$. We will label elements of
the
weight lattice with half-integers, so that the sub-lattice generated
by the root consists of the integers. With this labelling, the
integers and half-integers correspond to the two distinct basic
representations of the affine algebra $A_1^{(1)}$. To make a connexion
with the complex-ZS hierarchy one chooses a fixed element of the
weight lattice, that is a {\it half-integer\/} $n$. Then
$$
\psi={\ttau_{n+1}\over\ttau_n}\ ,\qquad
\pb={\ttau_{n-1}\over\ttau_n}\ ,
\nfr{psitau}
where we have used a tilde in order to avoid confusion with the
mKdV/KdV tau-functions. In addition, the equations of the hierarchy
imply
$$
\psi\pb=-\partial_x^2\,{\rm log}\,\ttau_n\ .
\efr
The hierarchies for different choices of $n$ are isomorphic.
\section{From Hierarchies to Matrix Models}
Given a hierarchy and its string equation, to obtain the field theory
describing a matrix model after a double-scaling limit, one must first
identify the partition function of the field theory with some variable
in the hierarchy.

If $x$ is the cosmological constant then the `specific heat' is
$F''=-\del^2_x\,\log\,{\cal Z}$. Clearly $-F''$ has a well defined
scaling
dimension, from a hierarchical point of view. So, in principle, one
can construct all the terms of the correct dimension from the
hierarchy, and then see which can be integrated twice with respect to
$x$. However, even this would not determine the normalization of the
`specific heat'. In the absence of any additional physical
requirements, one has to appeal to the matrix model. It transpires
that the partition function of each particular model is related to the
tau-function of the hierarchy in a simple way:
$$
{\cal Z}_{\rm ZS}=\ttau_n,\quad {\cal Z}_{\rm KdV}=\tau_0^2,\quad
{\cal Z}_{\rm mKdV}=\tau_0\tau_1\ ,
\nfr{part}
where the tau-functions for the hierarchies where introduced in
section \S\tausec.

With $x$ interpreted as the cosmological constant, the `specific heat'
is nothing but the correlation function $-\langle
PP\rangle=-\del_x^2\log\,{\cal Z}$, where the
operator which couples to the cosmological constant is conventionally
denoted $P$ and called the {\it puncture operator\/}.
Finally, the insertion of an operator in a correlation function is
given by the corresponding flow in the hierarchy, for example
\fr
\vev{{\cal O}_i PP} \ =\ {\del\phantom{t}\over \del t_i} \vev{PP}
\efr
where ${\cal O}_0=P$ and the right-hand side is computed using the
equation of the hierarchy. Moreover, with the identifications \part\
\fr
\vev{PP}_{\rm ZS}\ =\ -\psi\pb\ =\ \vev{PP}_{\rm NLS}\ ,\quad
\vev{PP}_{\rm KdV} \ =\ u\ ,\quad \vev{PP}_{\rm mKdV}\ =\ -\nu^2\ .
\efr

Let us consider first the hierarchies which should correspond to the
anti-hermitian matrix models and their correlation functions
constructed using the prescription above. The
anti-hermitian 1-matrix model in the 1-arc sector should correspond to
the sum of two KdV hierarchies in the variables $\chi=r-is$,
$\overline\chi =r+is=\chi^*$. Although these variables are complex
($r$ and $s$ are real functions of $x$), it turns out
that the correlation functions and the string equations are
real. Indeed, the `specific
heat' is given by the sum of the `specific heats',
\fr
\vev{PP}\ =\ \frac{1}{2} (\chi +\overline\chi)\ =\ r
\efr
and the correlation functions are given by
\fr
\vev{{\cal O}_k PP}\ =\ \frac12{\del_x} \left(
R_{k+1}(\chi) + R_{k+1} (\overline\chi) \right)\ .
\efr
Using the recursion relations of the KdV hierarchy it is easy to show
that $R_k(\overline\chi)=R_k(\chi)^*$ and hence the correlation
functions are real. Moreover,
the sum and the difference of the string equations \intstreq\ for
$\chi$ and $\overline\chi$ give
exactly eqs. \unoa\ for this model.

The anti-hermitian 1-matrix model in the 2-arc sector corresponds to
the ZS hierarchy. First of all, notice that for the ZS and NLS models
one has
\fr
\vev{{\cal O}_k P}\ =\ \frac{1}{2} H_{k+1}
\efr
implying, for example, $\vev{PP} = \frac{1}{2} H_1 = -\psi\pb$. For
the ZS hierarchy both $\psi$ and $\pb$ are real, and thus the whole
hierarchy and all the physical quantities are real. The string
equations \strineq\ correspond to the sum and difference of the two
equations \unoa\ where the second equation is multiplied by $i$.

We now turn our attention to the hermitian 1-matrix model. In the
1-arc sector we
have a double KdV hierarchy where everything is expressed in terms of
real functions. In the 2-arc sector, instead, we found only some of the
string equations of the NLS hierarchy. Indeed, for
the NLS hierarchy, one can easily show that $F_{2k}$, $H_{2k}$ and
$G_{2k+1}$ are purely imaginary, whereas $F_{2k+1}$, $H_{2k+1}$ and
$G_{2k}$ are real. Thus, although the `specific heat' is real, many
correlation functions are complex or pure imaginary. This obviously
forbids an interpretation of the {\it full\/} NLS hierarchy as a field
theory obtained after a double scaling limit of a hermitian 1-matrix
model in the 2-arc sector.
Taking the sum and the difference of the string equations \strineq\
and using the recursion relations of the hierarchy one gets
[\Ref{moore}]
\fr
\sum_{k=0}^{\infty}(k+1)t_k G_k\ =\ 0\ , \qquad\quad
\sum_{k=0}^{\infty}(k+1)t_k F_k\ =\ 0 \ .
\efr
The string equations \unoa\ should correspond to the multi-critical
points $t_0=x$, $t_n=\ $constant and $t_k=0$ otherwise. Thus,
if $n$ is even the first
equation is real and the second is pure imaginary and they correspond
to eq. \unoa\ where the second equation is multiplied by $i$. These
are the string equations we found in the first chapter. For $n$ odd
instead, both equations are complex and so they could not arise from
the direct study of the matrix model.
\chapter{Virasoro Constraints}
In [\Ref{vircostr}] it was shown, for the models
described by the KdV hierarchy, that the string
equation, along with the hierarchy equations, could be reformulated as
an infinite number of Virasoro-like constraints on the square-root of
the partition function
of the model. These constraints have a natural interpretation in terms
of the Schwinger-Dyson equations for the loops of the matrix model.
For the KdV model the square-root of the partition function is the
tau-function of the hierarchy. The fact that the Virasoro constraints
act on the tau-function of the hierarchy seems to be a universal
feature of all the models, as will become apparent.
First we briefly review the case
for the KdV hierarchy. Using \part, \utau\ and the string equation
\kdveq, and integrating twice we deduce
$$
\left(\sum_{k=1}^\infty(k+{\textstyle {1\over2}})t_k{\partial\ \ \
\over\partial
t_{k-1}}+ \frac18 t_0^2\right)\tau_0\ =\ 0\ .
\efr
Following [\Ref{vircostr}], we use the recursion relations of the
hierarchy \kdvr\ and
the relation between $\tau_0$ and $u$ in \utau, which together imply
$$
\partial_x^2{\partial\over\partial
t_{k+1}}\log\tau_0=\left(\frac{1}{2}\partial_x^3+2u\partial_x+u'\right)
\partial_x{\partial\over\partial t_k}\log\tau_0,
\nfr{rectau}
in order to express the $t_k$ derivative in terms of the $t_{k+1}$
derivative.
With this relation, one finds that $L_k\tau_0=0$ implies
$L_{k+1}\tau_0=0$, where the $L_k$ are written below.
All of the constants of integration encountered
are set to zero, on the grounds that they
would otherwise introduce spurious scales into the theory, except for
the constant in the
$L_0$ constraint which is dimensionless and
fixed by the requirement that the
algebra of constraints closes.
The end result is that the tau-function satisfies an
infinite number of constraints of the form
$$
L_{n}\tau_0=0,\quad\qquad n\geq-1\ ,
\efr
where the $L_n$'s are the Virasoro generators of a ${\bf Z}_2$-twisted
scalar field
$$
\eqalign{L_{-1}&=\sum_{m=1}^\infty(m+{\textstyle {1\over2}} )
t_m{\partial\ \ \over\partial t_{m-1}}+{\textstyle {1\over8}}t_0^2\cr
L_0&=\sum_{m=0}^\infty(m+{\textstyle
{1\over2}})t_m{\partial\ \over\partial t_m}+{\textstyle {1\over16}}\cr
L_n&=\sum_{m=0}^\infty(m+{\textstyle
{1\over2}})t_m{\partial\ \ \over\partial t_{m+n}}+{\textstyle
{1\over2}}\sum_{m=1}^n{\partial^2\ \ \ \over\partial t_{m-1}\partial
t_{n-m}}\ .\cr}
\nfr{virkdv}
\section{Virasoro Constraints for the MKDV Hierarchy}
\setchap{mkdvsec}
Although the matrix model which leads to the mKdV hierarchy has been
discussed in the literature [\Ref{mkdvmm}], the analogue of the
Virasoro constraints do not seem to be have been determined before
(although
ref. [\Ref{unimoore}] does discuss Virasoro constraints before
taking the double scaling limit). In this section we find the
constraints using the mKdV string equation \mkdveq\ and the recursion
relations for the hierarchy.

The mKdV string equation \mkdveq\ is obtained from integrating \nustr.
{}From \nustr\ one easily deduces
$$\sum_{k=0}^\infty(2k+1)t_k{\partial\nu^2\over\partial t_k}+2\nu^2
=0.
\nfr{nusqstr}
Recall that the partition function of the mKdV model is equal to the
product of the tau-functions $\tau_0$ and $\tau_1$.
We now express $\nu$ in \nustr\
in terms of the tau-functions $\tau_0$ and $\tau_1$, using \utau,
 and $\nu^2$ in \nusqstr\ using \tausq.
The resulting two equations can be decoupled to arrive at
$$
\del^2_x\left[\sum_{k=0}^\infty(2k+1)t_k{\partial\over\partial t_k}
\log\tau_j\right]=0\ ,\qquad j=0,1\ .
\efr
Integrating twice, and eliminating dimensionful
integration constants, the two resulting equations may be written
simply as
$$L_0\tau_j=\mu_j\tau_j\ ,\quad\qquad j=0,1\ ,
\nfr{hws}
where $L_0$ is identical to the Virasoro constraint of the KdV model,
eq. \virkdv, and $\mu_0$ and $\mu_1$ are two, {\it a priori\/}
undetermined, dimensionless integration constants. They are
not, however, independent as we now show. By substituting the
expression for $\nu$ in terms of the ratio $\tau_1/\tau_0$, in the
equation for the flows \dued, we deduce
$$
{\partial\over\partial t_k}\log\left({\tau_1\over\tau_0}\right)=
-\frac{1}{2}D^\star R_k.
\efr
This can now be substituted directly in \mkdveq\ to yield
$$
\sum_{k\geq0}(2k+1)t_k\left({1\over\tau_1}{\del\tau_1\over\del t_k}-
{1\over\tau_0}{\del\tau_0\over\del t_k}\right)=0.
\efr
The above, along with eq. \hws\ implies that $\mu_0=\mu_1\equiv\mu$.

Before we discuss
the possible meaning of the constant $\mu$, we first present a simple
argument for determining the higher Virasoro constraints.
We already know from the construction of the KdV constraints, that
$L_k\tau_0=0$ implies $L_{k+1}\tau_0=0$. It is also straightforward to
verify that if $L_0\tau_0=\mu\tau_0$ then $L_1\tau_0=0$, regardless of
the value of $\mu$. Therefore we deduce that $\tau_0$ satisfies the
infinite set of constraints
$$
L_n\tau_0=\mu\tau_0\,\delta_{n,0}\ ,\quad\qquad n\geq0\ .
\nfr{onecon}
To find the constraints satisfied by $\tau_1$, we notice that
$\tau_1$ satisfies exactly the same recursion relation as
$\tau_0$, that is \rectau, except that $u=2\partial_x^2\log\tau_0$ is
replaced
with $\tilde u=2\partial_x^2\log\tau_1$. Therefore, the same arguments
that were applied to determine the constraints on $\tau_0$ will be
applicable to $\tau_1$, hence we deduce
$$
L_n\tau_j=\mu\tau_j\,\delta_{n,0}\ ,\quad\qquad n\geq0\ ,
\nfr{mkdvvircon}
for $j=0$ and $1$.
So the mKdV partition function is the product of two factors which
separately satisfy a set of Virasoro constraints, however, in contrast
to the KdV case there is no $L_{-1}$ constraint and the $L_0$
constraint includes an undetermined constant. Notice that the
requirement that the constraints form a closed algebra under
commutation, does not in any way constrain the value of the constant.
It is important to realize that $\tau_0$ and $\tau_1$ are not
independent, in fact they satisfy a whole hierarchy of equations for
which
\reltau\ is but the first. So although, at first sight, the mKdV
Virasoro constraints look less restrictive than the KdV constraints,
one must bear in mind the additional equations which tie $\tau_0$ and
$\tau_1$ together.

The appearance of a parameter, which is not determined from the matrix
model, seems, at first sight, to be surprising. However, it is not
totally unexpected, indeed, such an occurrence is found at the first
critical point of the mKdV model. At this point, the square root
of the specific heat, $\nu$, satisfies the Painlev\'e II equation
[\Ref{mkdvmm}]
$$\nu''-2\nu^3+x\nu=0\ .
\nfr{painii}
This equation is known to admit a one-parameter family of solutions
[\Ref{math}]. The actual solution which describes
the matrix model, requires a scaling
behaviour $\nu\sim z^\xi$, as $x\rightarrow\infty$. Ref. [\Ref{watt}]
discusses how, for one particular value of the parameter, such a
physical solution does exists and is unique. It would be natural to
suggest
that the parameter is related to $\mu$, the eigenvalue of the $L_0$
constraint. Indeed, \painii\ admits the trivial solution $\nu=0$,
which corresponds to the situation when $\mu=\frac{1}{16}$, for which
the Virasoro constraints have the solution $\tau=1$ (i.e. $\tau$
being the vacuum of the twisted Fock space).
Notice that, for the KdV model, the Virasoro constraints are those of
an $sl(2,{\mmmath C})$ vacuum, whereas, for the mKdV model, the
Virasoro constraints are those of a highest weight state of conformal
dimension $\mu$.
\section{Virasoro Constraints for the ZS Hierarchy}
The analogous constraints for the ZS hierarchy and string equation
were found in [\Ref{moore}]. Here, we briefly repeat their derivation
which leads to Virasoro type constraints for an untwisted boson.

The string equations are (see eq. \strineq)
$$
\sum_{k\ge 0} (k+1) t_k F_k =0\ ,\quad \qquad
\sum_{k\ge 0} (k+1) t_k G_k =0\ .
\efr
Consider first the objects
$$
I_{j} = \sum_{k\ge 0} (k+1) t_{k} H_{k+j}' = \sum_{k\ge 0} (k+1) t_{k}
(gG_{k+j} - fF_{k+j})\ .
\efr
They can be reduced, using the hierarchy, to sums involving only
$F_{k}$, $G_{k}$ and their derivatives, which are related to the
string  equations.

{}From $I_0$, integrating twice over $x$ and introducing an arbitrary
integration
constant\note{We will discard all integration constants which would
have non-trivial dimension.} $\alpha$, we get
$$
\sum_{k\ge1} (k+1) t_k \langle {\cal O}_{k-1}\rangle +
{\alpha t_0 \over2} = 0
\efr
which will lead to the $L_{-1}$ constraint.

{}From $I_1$ we get
$$
\sum_{k\ge0} (k+1) t_k \langle {\cal O}_{k}\rangle + \beta = 0\ ,
\efr
where we have picked up a new dimensionless integration constant,
$\beta$. This leads to the $L_0$ constraint.

Using a similar procedure, from $I_2$ we get
$$
\sum_{k\ge0} (k+1) t_k \langle {\cal O}_{k+1}\rangle+ \alpha \langle P
\rangle = 0
\efr
which leads to the $L_1$ constraint.

Finally, with a few more steps from $I_3$ we obtain
$$
\sum_{k\ge0} (k+1) t_k \langle {\cal O}_{k+2}\rangle + \langle P
\rangle^2 +
\langle PP\rangle + \alpha \langle {\cal O}_1\rangle = 0\ .
\efr
Whilst the previous equations involve only first order derivatives
of the
partition function, the equation coming from $I_3$ has second order
terms,
which fix the function on which the Virasoro constraints act. In fact,
we can write
$$
\langle P\rangle^2 + \langle PP\rangle = (F')^2 - F'' = {1\over {\cal
Z}} \partial^2_x {\cal Z}
\efr
and, therefore, the Virasoro constraints act on
the partition function. Since for this model the partition function is
equal to the tau-function, we find that the Virasoro constraints act
on the tau-function, mirroring the situation for the KdV model.

By consistency, the commutator of two constraints should be a new
constraint on
the partition function. Therefore, using $[L_n,L_1]\equiv (n-1)L_{n+1}$
with $n\ge2$, we get an infinite set of constraints acting on the
partition function. These constraint are the Virasoro constraints
$$
L_n \tau^{\rm ZS}\ =\ 0\ , \qquad\qquad n\ge -1
\efr
where
$$
\eqalign{
L_{-1}&=\sum_{k\ge1}(k+1) t_k {\partial\over\partial t_{k-1}} + {\alpha
t_0\over2} \cr
L_{0}&=\sum_{k\ge0}(k+1) t_k {\partial\over\partial t_{k}} +
{\alpha^2\over4}\cr
L_{1}&=\sum_{k\ge0}(k+1) t_k {\partial\over\partial t_{k+1}} +
\alpha{\partial\over\partial t_0} \cr
L_{n}&=\sum_{k\ge0}(k+1) t_k {\partial\over\partial t_{k+n}}
+\sum_{k=0}^{n-2} {\partial^2\over \partial t_k \partial t_{n-k-2}} +
\alpha {\partial\over\partial t_{n-1}}\cr}
\nfr{vircst}
and $\beta=\alpha^2/4$ has been fixed by the relation $\lbrack
L_1,L_{-1}\rbrack = 2
(L_0 +\alpha^2/4 -\beta)=2L_0$.

\section{Connexion with the Tau-Function Formalism}

It was noticed in ref. [\Ref{moore}] that the Virasoro
constraints \vircst\ are those of an
untwisted scalar field. In the convention for which
$$
\varphi(z)=q-ip\,{\rm log}\, z+i\sum_{n\in{\bf Z}\neq0}a_n{z^{-n}\over
n}\ ,
\efr
and $[a_m,a_n]=n\delta_{n+m,0}$ and $[q,p]=i$, we have for $k\geq 0$
$$
t_k={\sqrt2\over k+1}a_{-k-1},\qquad \quad {\partial\over\partial t_k}
= {1\over\sqrt2}a_{k+1}\ .
\efr
The zero-mode $p$ is related to the integration constant $\alpha$ via
$p=\alpha/\sqrt2$. The conjugate variable to $p$ does not
appear in the Virasoro operators.

The fact that the Virasoro constraints are those of an untwisted
scalar field is very natural from the point of view of the
tau-function approach, which we explained in section \S\tausec\ based
on ref. [\Ref{kacwak}].
Recall that the zero mode of the scalar field of the construction of
ref. [\Ref{kacwak}] must
have the quantized values $m/\sqrt2$, for $m\in{\mmmath Z}$, in order
that the vertex operators have local expansions. The tau-function can
be projected onto eigenspaces of the zero-mode, these are precisely
the $\ttau_n$ which where introduced in \S\tausec, with $\sqrt2 n$
being the eigenvalue of the zero-mode.

{}From eq. \part\ the partition function of the ZS model is
equal to $\ttau_n$, for some fixed half-integer $n$. It is very
natural to identify the scalar field of the Virasoro constraints with
the scalar field of the Hirota equations of [\Ref{kacwak}]. We do not
yet have a direct proof of this, however, below we present some
arguments which support this view.
Given this identification, one is led to a relation between the
parameter
$n$ and the integration constant of the Virasoro constraints $\alpha$:
$$
p\ =\ {\alpha\over\sqrt2}\ =\ -\sqrt2 n,\qquad
n\in\frac12{\mmmath Z}\ .
\nfr{alfaeq}
The possibility that $\alpha$ is quantized seems to be consistent with
the
results that we obtain in \S3.4 and \S3.5, for the KdV and mKdV
reductions which require $\alpha=0$ and $\alpha=-1$,
respectively\note{However, the `topological' point described in ref.
[\Ref{moore}] does not seem to require any particular value of
$\alpha$.}.
To substantiate this identification we now show that if $\ttau_n$
satisfies
the Virasoro constraints eq. \vircst\ with $\alpha$, then
$\tau_{n\pm1}$ satisfy the same constraints but with $\alpha\rightarrow
\alpha\mp 2$. We prove this fact following a similar
demonstration as the previous paragraph. Let us consider the objects
$$
Y_{j} = \sum_{k\ge -1} (k+1) t_{k}{\partial\ \ \over\partial t_{k+j-1}}
\left(\ttau_{n\pm1}\over\ttau_n\right)\ .
\efr
and use the string equations, the hierarchy of Hirota equations for
the $\ttau_n$'s, and the fact that $\ttau_n$
satisfies Virasoro constraints with $\alpha$, $L_m \ttau_n \equiv L_m
(\alpha)\ttau_n =0$, $m\ge-1$. Considering $Y_0$, $Y_1$ and $Y_2$,
it is easy to obtain
$$
L_{-1}(\alpha\mp2)\ttau_{n\pm1}=L_{0}(\alpha\mp2)\ttau_{n\pm1}=
L_{1}(\alpha\mp2)\ttau_{n\pm1}=0\ .
\efr
Again, the $L_2$ constraint is more tricky. Considering $Y_3$, it is
easy to show that
$$
L_2(\alpha\mp2) \ttau_{n\pm1} = (\alpha\mp2)\left\{ \ttau_n \left(
{\partial\over\partial t_1} \mp {\partial^2\over\partial t_{0}^{2}
}\right)
\left(\ttau_{n\pm1}\over\ttau_n\right) \mp 2 \ttau_{n\pm1} \partial^2_x
{\rm log}\, \ttau_n\right\}
\efr
The right hand side of this equation vanishes because of one of the
Hirota
equations satisfied by the tau-functions. In particular, (see ref.
[\Ref{kacwak}] pg. 232 $(III)_{1;n,n+1}$),
$$
L_2(\alpha\mp2) \ttau_{n\pm1}=(\alpha\mp2){1\over\ttau_n}
(D_{1}^{2} \mp D_2 )\ttau_{n\pm1}\cdot\ttau_n =0\ .
\efr
The $D_i$'s are operators of the Hirota calculus which are defined in
ref. [\Ref{kacwak}] for example. Therefore, we get
$$
L_m(\alpha) \ttau_n = L_m(\alpha\mp2) \ttau_{n\pm1}=0\ ,\qquad m\ge
-1\ .
\nfr{shiftvir}
If we now consider $p=\alpha/\sqrt2$ as the zero-mode, `momentum'
operator
of the scalar field, with $p\,\ttau_n=-\sqrt2 n\ttau_n$, in accordance
with \alfaeq, then its conjugate variable or `position' operator is
$q=-it_{-1}/\sqrt2$. This is deduced from equation \tmodep\ and
\shiftvir.

The above result \shiftvir\ also implies that the whole Hirota
hierarchy
admits a `master' string equation. It is most suggestively written in
terms of the full tau-function $\ttau$, for which the $\ttau_n$ are the
projections onto eigenspaces of the zero-mode. The master
string equation is the $m=-1$ version of the the following Virasoro
constraints
$$L_m\ttau=0,\ \ \ m\geq-1\ ,$$
where the $L_m$ are the Virasoro generators of the bosonic field. So
the final set of constraints are exactly analogous to those for the
1-arc KdV case, the difference being that there one has a twisted
scalar field, whereas here we have an untwisted scalar field.

The appearance of the parameter $\alpha$ is rather mysterious, since it
was not manifest in the matrix model. It seems to label
different sectors in the theory which are not connected by the flows.
It is clearly desirable to have a better understanding of its origin
and meaning.
\section{KDV Reduction of Virasoro Constraints}
We have already shown how the the ZS hierarchy can be reduced to the
KdV
hierarchy, and how the string equations respect the reduction. On
general grounds, one would anticipate that this would extend to all
the Virasoro constraints, and we now prove this. The reduction
involves taking $\pb=-e^{t_{-1}}$ and $t_{2k+1}\rightarrow0$ $\forall
k$. Then $u=-\psi\pb$ satisfies the KdV hierarchy with
$$
t_{2k}^{\rm ZS}\equiv 2^{-k} t_{k}^{\rm KdV},\quad\qquad
\tau^{\rm ZS}=\left(\tau^{\rm KdV}\right)^2
\nfr{blob}
and
$$
{\partial u \over\partial t_{k}^{\rm KdV}} = \partial_x R_{k+1}\ .
\efr
Notice that the second equation of \blob\ implies that under the
reduction, i.e. on the subspace $t_{2k+1}=0$ $\forall k$ with
$\pb=-e^{t_{-1}}$, ${\cal
Z}_{\rm ZS}\rightarrow{\cal Z}_{\rm KdV}$, as it should.
Under such reduction, $H_{2k}=-H_{2k-1}'$:
$$
\eqalign{
F_{2k} + G_{2k}& = (F_{2k-1} +G_{2k-1})' + (g+f) H_{2k-1} = (g-f)
H_{2k-1}\cr
0&=F_{2k+1} + G_{2k+1} = (F_{2k} +G_{2k})' + (g+f) H_{2k}\cr
&\quad\quad\Rightarrow \left((g+f)H_{2k-1}\right)' + (g+f)H_{2k}=0\cr}
\efr
but $(g+f)'=2\pb\,'=0$, and the result follows.

We now show that the ZS Virasoro constraints on $\tau^{\rm
ZS}(=\ttau_n)$ reduce to
Virasoro constraints on $\tau^{\rm KdV}(=\tau_0)$ for a precise value
of the
zero-mode $\alpha$ (or $n$). Let us
first consider the equation corresponding to $I_0$, under the reduction
$$
\sum_{k\ge1}(2k+1)t_{2k} H_{2k}+\alpha=-\sum_{k\ge1}(2k+1)t_{2k}
H_{2k-1}' +\alpha=0
\efr
Using the relations with the correlation functions, and integrating
twice over $t_0=x$, we get
$$
\sum_{k\ge1}(2k+1)t_{2k} \langle {\cal O}_{2k-2}\rangle - {\alpha
t_{0}^{2} \over4}=0
\efr
which will produce the reduced $L_{-1}$ constraint. The reduction
of $I_1$ is direct, and leads to
$$
\sum_{k\ge0}(2k+1)t_{2k}\langle {\cal O}_{2k}\rangle +{\alpha^2
\over4}=0
\nfr{duee}
which produces the $L_0$ constraint. This constraint is also obtained
from $I_2$, the $L_1$ constraint in the ZS hierarchy. The
corresponding equation is
$$
\eqalign{
&\sum_{k\ge0} (k+1) t_k H_{k+2}' + \alpha H_{1}' + 2 H_2 =0 \cr
&\Rightarrow \sum_{k\ge0} (2k+1)t_{2k} H_{2k+1}'' + (2-\alpha) H_{1}'=0
\cr}
\efr
Using the relations with the correlation functions, and integrating
three times over $t_0=x$, we get
$$
\partial^3_x \left[ \sum_{k\ge0}(2k+1)t_{2k}\langle {\cal O}_{2k}
\rangle + (1+\alpha) F \right]\ =\ 0
\efr
which, by consistency with eq. \duee, requires $\alpha=-1$.
Therefore, the KdV reduction is only consistent for this value of
the, {\it a priori}, arbitrary parameter $\alpha$.

Let us now consider the equation corresponding to $I_3$, which, again,
will to fix the function on which the constraints act:
$$
\sum_{k\ge0} (2k+1) t_{2k} H_{2k+3}'+4\langle PPP\rangle\langle P
\rangle + 4\langle PP\rangle^2  + 2H_3
+2(1-\alpha)\langle PPPP\rangle =0\ .
\efr
In the usual way, we get
$$
\sum_{k\ge0} (2k+1) t_{2k} \langle {\cal O}_{2k+2}\rangle + \langle
P\rangle^2 + (1-\alpha)\langle PP\rangle =0\ .
\efr
This equation will lead to the $L_1$ constraint, and, again, the
second order
terms fix  the functional on which the Virasoro constraint act. In
this case $\alpha=-1$, and
$$
\langle P\rangle^2 + (1-\alpha)\langle PP\rangle=\langle P\rangle^2 + 2
\langle PP\rangle ={4\over\sqrt{{\cal Z}_{\rm KdV}}}\partial^2_x
\sqrt{{\cal Z}_{\rm KdV}} \equiv {4\over\tau^{\rm KdV}}\partial^2_x
\tau^{\rm KdV}\ .
\efr
Notice that the required value of $\alpha$ is consistent with the
quantization proposed in \alfaeq.
Therefore, the reduced Virasoro constraints act on the square root of
the
partition function, which is the tau-function of the KdV hierarchy, in
agreement with [\Ref{vircostr}].

In terms of the variables $t_k \equiv t_{k}^{\rm KdV} = 2^{k}
t_{2k}^{\rm ZS}$, the reduced Virasoro constraints are
$\tilde L_n \tau^{\rm KdV} =0$,
with $n\ge-1$, where the operators $\tilde L_n$ are those of eq.
\virkdv.
\section{MKDV Reduction}
In the case of the mKdV reduction we have already shown in section
\S\redsec\  that $G_{2k+1}$,
$F_{2k}$ and $H_{2k}$ vanish. It is straightforward to verify that
under the reduction, i.e. on the subspace $t_{2k+1}=0$ $\forall k$
with
$\pb=e^{2t_{-1}}\psi$, ${\cal Z}_{\rm ZS}\rightarrow{\cal Z}_{\rm
mKdV}$.
Using these results one can apply the reduction directly on the
constraints.

{}From $I_0$ ($L_{-1}$ constraint), we get
$$
\sum_{k\ge1}(2k+1)t_{2k} H_{2k} +\alpha=0\quad \Rightarrow \qquad
\alpha=0\ .
\efr
Therefore, the mKdV reduction requires $\alpha=0$, which is clearly
consistent with the quantization of $\alpha$ proposed in \alfaeq.
{}From $I_1$, we get
$$
\sum_{k\ge0}(2k+1)t_{2k} H_{2k+1}' + 2H_1=0\ .
\efr
But,
$$
H_{2k+1}'=2\langle PP{\cal O}_{2k}\rangle\equiv -2 {\partial F''
\over\partial t_{2k}}
\efr
and we get the equation
$$
\sum_{k\ge0}(2k+1)t_{2k} {\partial F'' \over\partial t_{2k}} +2F''=0
\efr
which has the form of an $L_0$ constraint. This equation can also be
rewritten as
$$
\sum_{k\ge0}(2k+1)t_{2k} {\partial \nu \over\partial t_{2k}} +\nu=0
\efr
which, after integration, becomes the string equation of the mKdV
hierarchy, eq. \mkdveq.

If we reduce the $L_0$ constraint of the ZS hierarchy itself, we find
that the partition of the mKdV model satisfies
$$
\sum_{k\geq0}(2k+1)t_k^{\rm mKdV}{\partial\ \ \ \ \over\partial
t_k^{\rm
mKdV}}{\cal Z}_{\rm mKdV}=0\ .
\efr
Notice that $\mu$, the parameter of the mKdV
Virasoro constraints of \S\mkdvsec, is determined by the reduction to
be $\frac{1}{16}$.

In a similar way, from $I_{2k}$ we get equations which are identically
zero and do not give rise to any constraint. Instead, from $I_3$
($L_{2}$ constraint) we get
\fr
\left(\sum_{k\ge0}(2k+1)t_k^{\rm mKdV}
{\partial \ \ \ \  \over\partial t_{k+1}^{\rm mKdV}} +
\frac{1}{2}{\del^2\ \over\del
x^2}\right) {\cal Z}_{\rm mKdV} \ = \ 0
\nfr{chi}
where $t^{\rm mKdV}_k=2^kt^{\rm ZS}_{2k}$. We can write the above
constraint in the following way. Firstly, we express the partition
function in terms of the tau-functions ${\cal Z}_{\rm
mKdV}=\tau_0\tau_1$. Then we use the relation \reltau\ to write
$$\partial_x^2(\tau_0\tau_1)=2\tau_0''\tau_1+2\tau_0\tau_1'',
\efr
from which we deduce that \chi\ may be rewritten as
$$
\left(L_1\tau_0\right)\tau_1+\tau_0\left(L_1\tau_1\right)=0\ ,
\efr
which is clearly a consequence of the $L_1$ constraints for $\tau_0$
and $\tau_1$ that we found for the mKdV model in \S\mkdvsec.

One could carry on this process of reducing the higher Virasoro
constraints. The resulting constraints would act directly on the
partition function of the mKdV model; and hence would not be Virasoro
constraints.
Nevertheless, we expect that the constraints
on the partition function should be expressible in terms of Virasoro
constraints acting on  each tau-function separately, as we found for
above for $L_1$. In any case,
one would always find an expression which was compatible with
the mKdV Virasoro constraints: this is due to the fact that the mKdV
Virasoro constraints follow directly from the mKdV string equation and
the
recursion relations of the hierarchy, which are obtained by the
reduction
from the ZS string equation and hierarchy.
\chapter{Discussion and Open Problems}
In this paper we have attempted to analyse all the possible double
scaling
limits of the hermitian and anti-hermitian 1-matrix models. As it is
clear from the fact that eq. \unop, after the double scaling limit,
gives rise to a differential
operator of degree two in $x$, the hermitian and anti-hermitian
1-matrix models are related to the $sl(2,{\mmmath C})$ hierarchies.
Hermitian and anti-hermitian matrix models have many common
properties: in the 1-arc sector they both give rise to a KdV
hierarchy for an even potential and to a doubled KdV hierarchy
for a general potential; in the 2-arc sector, with even
potential, they both give rise to the mKdV
hierarchy. Instead they differ in the 2-arc sector with a general
potential where the hermitian models give rise to only half of the
critical points associated to the NLS hierarchy whereas the
anti-hermitian model gives all the critical points associated to the
ZS hierarchy (except for the `Topological' one).

For the hermitian models the multi-critical points obtained, although
described by a NLS hierarchy, actually have solutions which are
described by a mKdV hierarchy. We do not know whether these solutions
are the only ones. Furthermore, these critical points correspond to
purely even potentials; in this sense the NLS structure is irrelevant,
and one is really dealing with a mKdV structure.
Instead the ZS hierarchy admits also a reduction to KdV, in other
words in the 2-arc sector of the anti-hermitian matrix model we found
a {\it new\/} series of multi-critical points described by the KdV
hierarchy,
besides the one already known from the
1-arc sector. This set of KdV multi-critical points are
not in any simple way connected with those in the 1-arc
sector, since, for example, the topological critical point describing
topological gravity [\Ref{topwitten}], which
cannot be obtained from the 1-arc sector, is obtained from the 2-arc
sector with anti-hermitian matrices with a fourth order potential.

The situation in the 2-arc sector with anti-hermitian matrices and a
general potential seems to be the richest, being described by the ZS
hierarchy. The `even' multi-critical
points admit solutions described by KdV and mKdV hierarchies, which
require two particular values of the parameter $\alpha$.
Clearly, it would desirable to
understand the r\^ole of the parameter $\alpha$, from the point of
view
of the matrix model, and also to know whether the KdV and mKdV
solutions exhaust the possible solutions for the `even' multi-critical
points.
For instance, are there other solutions for different values of
$\alpha$, and do solutions exist only for the discrete values
suggested by the tau-function formalism?
An interesting open question regards the nature of the
`odd' multi-critical points of the ZS string equation.
It is now well-known that solutions for the KdV and mKdV systems
describing multi-critical behaviour exist
[\Ref{oldmat},\Ref{unimoore},\Ref{watt}]; we do not have any arguments
to show that solutions can be found for the `odd' scaling points of the
ZS
hierarchy, except for the first (or `topological') point, corresponding
to
$t_1\neq0$, which was investigated in [\Ref{moore}]. This point cannot
be
obtained from the matrix model and, as described in ref.
[\Ref{moore}], could give rise to a new kind of `topological' theory.
Anyway, since the mathematical apparatus exists for
tackling the issue of the existence of solutions to these non-linear
differential equations [\Ref{bigmoore}], we hope these questions will
be addressed and solved elsewhere.

One of the results of this paper is the realization of the rather
universal nature of the Virasoro constraints and the fact that they
act on the tau-functions of the appropriate hierarchy, and not,
necessarily, directly on the partition function. For the KdV and ZS
hierarchies, the Virasoro constraints are
$$
L_n\tau=0\qquad\quad n\geq-1\ ,
\efr
where in both cases $\tau$ is an element of the
basic representation of the Kac-Moody algebra $A_1^{(1)}$, for the
twisted and untwisted constructions, respectively. In the mKdV case
the partition function is the product of two tau-functions, which
arise from the two basic representations of $A_1^{(1)}$, which both
satisfy the Virasoro constraints of a highest weight vector:
$$
L_n\tau_j=\mu\tau_j\,\delta_{n,0},
\efr
for $j=0$ and $1$.

Near the completion of this work our attention was drawn to
refs. [\Ref{new},\Ref{dalley}]  which discuss similar Virasoro
constraints to those above.
\acknowledgments
We would like to thank \v C.~Crnkovi\'c, M.~Douglas and G.~Moore for
having sent us the preprint [\Ref{moore}] prior to publication. We
would like to thank M. Newman for explaining his recent work with D.
Gross [\Ref{new}]. The research of T.H. is supported by NSF
PHY90--21984, that of J.M. by a Fullbright/MEC fellowship and
that of A.P. by an INFN fellowship. C.N. is partially supported by the
Ambrose Monell Foundation.
\references
\beginref
\Rref{unimoore}{\v C.~Crnkovi\'c, M.~Douglas and G.~Moore, ``{\sl
Physical
solutions for unitary-matrix models\/}", Nucl. Phys. {\bf B360} (1991)
507.}
\Rref{moore}{\v C.~Crnkovi\'c, M.~Douglas and G.~Moore, ``{\sl Loop
equations and the topological phase of multi-cut matrix models\/}",
preprint YCTP-P25-91, RU-91-36, August 1991.}
\Rref{GD}{I.M.~Gel'fand and L.A.~Dikii, ``{\sl Asymptotic behaviour of
the
resolvent of Sturm-Louville equations and the algebra of the
Korteweg-de Vries equation\/}", Russ. Math. Surv. {\bf30} (1975)
77; ``{\sl Fractional Powers of Operators and Hamiltonian Systems\/}",
Funkts. Anal. Pril. {\bf10} (1976) 13.}
\Rref{vircostr}{R. Dijkgraaf, E. Verlinde and H. Verlinde, ``{\sl Loop
equations and Virasoro constraints in non perturbative 2d quantum
gravity\/}", Nucl. Phys. {\bf B348} (1991) 435.\newline
M. Fukuma, H. Kawai and R. Nakayama,
``{\sl Continuum Schwinger-{\goodbreak}Dyson
equations and universal structures in 2-dimensional quantum
gravity\/}", Int. J. Mod. Phys. {\bf A6} (1991) 1385.}
\Rref{kacwak}{V.G.~Kac and M.~Wakimoto, ``{\sl Exceptional Hierarchies
of Soliton Equations\/}", Proceedings of Symposia in
Pure Mathematics {\bf49} (1989) 191.}
\Rref{wilson}{B.A.~Kupershmidt and G.~Wilson, ``{\sl Modifying Lax
Equations and the Second Hamiltonian Structure\/}", Invent. Math.
{\bf 62} (1981) 403.}
\Rref{bigmoore}{G.~Moore, ``{\sl Geometry of the string equations\/}",
Comm. Math. Phys. {\bf133} (1990) 261; ``{\sl Matrix models of 2d
gravity and isomonodromic deformations\/}", in Common Trends in
Mathematics and Quantum Field Theories, Prog. Theor. Phys. Suppl.
{\bf 102} (1990) 255.}
\Rref{topwitten}{E. Witten, ``{\sl On the Structure of the Topological
Phase of two dimensional Gravity\/}", Nucl. Phys. {\bf B340} (1990)
281.\newline
R.~Dijkgraff and E.~Witten, ``{\sl Mean field theory, topological
field theory and multi-matrix models\/}", Nucl. Phys. {\bf B342}
(1990) 486. \newline
J.~Distler, ``{\sl 2d quantum gravity, topological field theory and
multi-critical matrix models\/}", Nucl. Phys. {\bf B342} (1990) 523.}
\Rref{penner}{R.C.~Penner, ``{\sl Perturbative series and the moduli
space of Riemann surfaces\/}", J. Diff. Geom. {\bf27} (1988) 35.
\newline
J.~Harer and D.~Zagier, ``{\sl The Euler characteristic of the moduli
space of curves\/}", Invent. Math. {\bf 85} (1986) 457.
\newline J.~Distler and C.~Vafa, ``{\sl A critical
matrix model at c=1\/}", Mod. Phys. Lett. {\bf A6} (1991) 259.\newline
C.~Itzykson and J.-B. Zuber, ``{\sl Matrix integration and the
combinatorics of the modular group\/}", Comm. Math. Phys. {\bf134}
(1990) 197.}
\Rref{konts}{M.~Kontsevich, ``{\sl Intersection theory on the space of
curve moduli\/}",  preprint 1990.\newline
E.~Witten, ``{\sl On the Kontsevich model and other models of two
dimensional gravity\/}", preprint IASSNS-HEP-91/24, June, 1991.}
\Rref{mkdvmm}{\v C. Crnkovi\'c and G. Moore, ``{\sl Multi-critical
multi-cut matrix models\/}", Phys. Lett. {\bf257B} (1991) 322.}
\Rref{douglas}{M. Douglas, {\sl Strings in less than one dimension and
the generalized KdV hierarchies\/}", Phys. Lett. {\bf238B} (1990) 176.}
\Rref{petropoulos}{C.~Bachas and P.~Petropoulos, ``{\sl Doubling of
equations and universality in matrix models of random surfaces\/}",
Phys. Lett. {\bf 247B} (1990) 363.}
\Rref{twoarc}{M.~Douglas, N.~Seiberg and S.~Shenker, ``{\sl Flow and
instability in quantum gravity\/}", Phys. Lett. {\bf B244} (1990)
381.\newline
P.~Mathieu and D.~S\'en\'echal, ``{\sl A well-defined multi-critical
series in hermitian one matrix models\/}", preprint LAVAL-PHY-21/91,
February 1991.}
\Rref{hier}{See for example: A.~Das, ``{\sl Integrable Models\/}",
Lectures Notes in Physics, Vol. 30, World Scientific, Singapore.}
\Rref{nappi}{C.~Nappi, ``{\sl Painlev\'e II and odd polynomials\/}",
Mod. Phys. Lett. {\bf A5} (1990) 2773.\newline
P.~Petropoulos, ``{\sl Doubling versus non-doubling of equations and
phase space structure in the one-hermitian-matrix models\/}", Phys.
Lett. {\bf 261B} (1991) 402.}
\Rref{periwal}{V.~Periwal and D.~Shevitz, ``{\sl Unitary-matrix models
as exactly solvable string theories\/}", Phys. Rev. Lett. {\bf 64}
(1990) 1326; ``{\sl Exactly solvable unitary matrix models:
multi-critical potentials and correlations\/}", Nucl. Phys. {\bf B344}
(1990) 731.}
\Rref{drinfel}{V.G.~Drinfel'd and V.V.~Sokolov, ``{\sl Lie algebras and
equations of Korteweg-de Vries type\/}", J. Sov. Math. {\bf30}
(1985) 1975.}
\Rref{zaksab}{V.E.~Zakharov and A.B.~Shabat, ``{\sl A scheme of
integrating non-linear equations of mathematical physics by the method
of the inverse scattering problem\/}", Funkts. Anal. Pril. {\bf8}
No. 3 (1974) 54.}
\Rref{watt}{A. Watterstam, ``{\sl A solution to the string equation of
unitary matrix models\/}", Phys. Lett. {\bf263B} (1991) 51.}
\Rref{oldmat}{F.~David, ``{\sl Planar diagrams, two-dimensional
lattice gravity and surface models\/}", Nucl. Phys. {\bf B257 [FS14]}
(1985) 45.\newline
V.A.~Kazakov, I.K.~Kostov and A.A.~Migdal, ``{\sl Critical properties
of randomly triangulated planar random surfaces\/}", Phys. Lett.
{\bf157B} (1985) 295.\newline
I.K.~Kostov and M.L.~Mehta, ``{\sl Random surfaces of arbitrary genus:
exact results for d=0 and -2 dimensions\/}", Phys. Lett. {\bf 189B}
(1987) 118.\newline
M.~Douglas and S.~Shenker, ``{\sl Strings in less than one
dimension\/}", Nucl. Phys. {\bf B355} (1990) 635.\newline
D.~Gross and A.~Migdal, ``{\sl A non-perturbative treatment of
two-dimensional quantum gravity\/}", Nucl. Phys. {\bf B340} (1990)
333.\newline
E. Brezin and V.A. Kazakov, ``{\sl Exactly solvable field theories of
closed strings\/}", Phys. Lett. {\bf 236B} (1990) 144.}
\Rref{witten}{E.~Witten, ``{\sl Two dimensional gravity and
intersection theory on moduli space\/}", preprint
IASSNS-HEP-90/45, May, 1990.}
\Rref{math}{S.P.~Hastings and J.B. McLeod, ``{\sl A boundary value
problem associated with the second Painlev\'e transcendent and the
Korteweg-de Vries equation\/}", Arch. Rat. Mech. and Anal. {\bf73}
(1980) 31.}
\Rref{new}{D.J. Gross and M. Newman, ``{\sl Unitary and Hermitian
matrices in an external field (2): the Kontsevich model and continuum
Virasoro constraints\/}", preprint PUPT-1282, September 1991.}
\Rref{morris}{S. Dalley, C. Johnson and T. Morris, ``{\sl
Multicritical complex matrix models and non-perturbative 2d quantum
gravity\/}", preprint SHEP 90/91-16, February 1991.}
\Rref{dalley}{S. Dalley, C. Johnson and T. Morris, ``{\sl
Non-perturbative two-dimensional Quantum Gravity\/}", preprint SHEP
90/91-28, June 1991.}
\endref
\ciao
